\documentclass[fleqn,usenatbib]{aa}  

\usepackage{graphicx}
\usepackage{txfonts}
\usepackage{newtxtext,newtxmath}
\usepackage[T1]{fontenc}
\usepackage{upgreek}
\usepackage{placeins}

\DeclareRobustCommand{\VAN}[3]{#2}
\let\VANthebibliography\thebibliography
\def\thebibliography{\DeclareRobustCommand{\VAN}[3]{##3}\VANthebibliography}

\defcitealias{4FGL}{4FGL}
\defcitealias{4FGL-DR3}{4FGL DR3}
\defcitealias{YMW16}{YMW16}
\defcitealias{3PC}{3PC}

\usepackage{lineno}

\usepackage{soul} 

\begin{document}

   \title{Radio and gamma-ray timing of TRAPUM L-band \textit{Fermi} pulsar survey discoveries}

\author{
M.~Burgay\inst{1}
\and
L.~Nieder\inst{2,3}
\and
C.~J.~Clark\inst{2,3}
\and
P.~C.~C.~Freire\inst{4}
\and
S.~Buchner\inst{5}
\and
T.~Thongmeearkom\inst{6,7}
\and
J.~D.~Turner\inst{6}
\and
E.~Carli\inst{6}
\and
I.~Cognard\inst{8,9}
\and
J.-M.~Grie{\ss}meier\inst{8,9}
\and
R.~Karuppusamy\inst{4}
\and
M.~C.~i~Bernadich\inst{4,1}
\and
A.~Possenti\inst{1}
\and
V.~Venkatraman~Krishnan\inst{4}
\and
R.~P.~Breton\inst{6}
\and
E.~D.~Barr\inst{4}
\and
B.~W.~Stappers\inst{6}
\and
M.~Kramer\inst{4,6}
\and
L.~Levin\inst{6}
\and
S.~M.~Ransom\inst{10}
\and
P.~V.~Padmanabh\inst{2,3}
}
\institute{
INAF -- Osservatorio Astronomico di Cagliari, Via della Scienza 5, I-09047 Selargius (CA), Italy\and
Max Planck Institute for Gravitational Physics (Albert Einstein Institute), D-30167 Hannover, Germany\and
Leibniz Universit\"{a}t Hannover, D-30167 Hannover, Germany\and
Max-Planck-Institut f\"{u}r Radioastronomie, Auf dem H\"{u}gel 69, D-53121 Bonn, Germany\and
South African Radio Astronomy Observatory, 2 Fir Street, Black River Park, Observatory 7925, South Africa\and
Jodrell Bank Centre for Astrophysics, Department of Physics and Astronomy, The University of Manchester, Manchester M13 9PL, UK\and
National Astronomical Research Institute of Thailand, Don Kaeo, Mae Rim, Chiang Mai 50180, Thailand\and
LPC2E - Universit\'{e} d'Orl\'{e}ans /  CNRS, 45071 Orl\'{e}ans cedex 2, France\and
Observatoire Radioastronomique de Nan\c{c}ay (ORN), Observatoire de Paris, Universit\'{e} PSL, Universit\'{e} d'Orl\'{e}ans, CNRS, 18330 Nan\c{c}ay, France\and 
National Radio Astronomy Observatory, 520 Edgemont Rd., Charlottesville, VA 22903 USA
}

   \date{Received 16/07/2024; accepted 26/09/2024}

 
  \abstract{This paper presents the results of a joint radio and gamma-ray timing campaign on the nine millisecond pulsars (MSPs) discovered as part of the L-band targeted survey of \textit{Fermi}-LAT sources performed in the context of the Transients and Pulsars with MeerKAT (TRAPUM) Large Survey Project. Out of these pulsars, eight are members of binary systems; of these eight, two exhibit extended eclipses of the radio emission. Using an initial radio timing solution, pulsations were found in the gamma rays for six of the targets. For these sources, a joint timing analysis of radio times of arrival and gamma-ray photons was performed, using a newly developed code that optimises the parameters through a Markov chain Monte Carlo (MCMC) technique. This approach has allowed us to precisely measure both the short- and long-term timing parameters. This study includes a proper motion measurement for four pulsars, which a gamma ray-only analysis would not have been sensitive to, despite the 15-year span of \textit{Fermi} data.}


   \maketitle
%

\section{Introduction}

Pulsars and, in particular, the fastest rotating millisecond pulsars (MSPs), are excellent laboratories for astrophysics and fundamental physics. Thanks to their clock-like nature, for instance, they can be used to test the predictions of relativistic gravity theories \citep[e.g.][]{FreireWex2024, Kramer2021_DPSR}, or as detectors of gravitational waves in the nano-hertz regime \citep[e.g.][]{Hobbs2010+PTA, Agazie2023_NanoGrav, Antoniadis2023_EPTA, Reardon2023_PPTA, gammaPTA}, or as probes of the equation of state (EoS) of nuclear matter \citep[e.g.][]{Hu2020_EoS} and other phenomena \citep{PSRHandbook}. 

Over one quarter of the presently known population of Galactic MSPs (see the ATNF Pulsar Catalogue PSRCat\footnote{\label{PSRcatfootnote}\url{https://www.atnf.csiro.au/people/pulsar/psrcat/}}, \citealt{psrcat}, and the GalacticMSP catalogue\footnote{\label{GalMSPsfootnote}\url{http://astro.phys.wvu.edu/GalacticMSPs/}}, including unpublished discoveries) have been discovered targeting unassociated gamma-ray point sources detected by the Large Area Telescope (LAT) instrument on board the \textit{Fermi} satellite \citep[][hereafter \citetalias{3PC}]{3PC}. These include some of the most extreme and/or elusive pulsars, such as the second tightest orbit binary MSP J1653$-$0158 \citep{Nieder2020+J1653}, which was only recently surpassed by J1953$+$1844, discovered in the M71 globular cluster \citep{Pan2023}; the fastest spinning MSP in the Galactic field \citep{Bassa2017+J0952}; and over two-thirds of all Galactic eclipsing binaries, namely, the so-called `spider' pulsars \citep{Roberts2013+Spiders}. The latter are categorised as either `black widows' (BWs), with short duration eclipses and companion stars with masses below 0.1 $M_\odot$, or `redbacks' (RBs), with eclipses covering large fractions of the orbital period and larger companion masses. Some RBs, namely, the so-called transitional MSPs, represent the missing link between accreting low-mass X-ray binaries and fully recycled MSPs (e.g. \citealt{Archibald2009Sci_J1023}).

The full scientific potential of MSPs as science probes can only be extracted through long-term follow-up campaigns aimed at finding a `coherent timing solution' that would provide very-high-accuracy astrometric, spin and (in the case of binary systems) orbital parameters. Typically, at least one year is needed to get an accurate position and ensure no covariance with the spin-period derivative measurement (a crucial parameter  in estimating pulsar energetics). Typically, several years are needed to get a proper motion measurement.

In the case of \textit{Fermi}-selected targets, once an initial timing solution is obtained from multiple radio observations, if high-energy pulsations are found and the signal-to-noise ratio (S/N) is high enough, gamma-ray photons can be used to extend the timing solution to cover the total time span of the \textit{Fermi} mission (15 years at the time of writing). This enables the immediate measurement of timing parameters that require longer time spans, minimising the required radio follow-up time. Timing solutions covering long time spans also have the advantage of making multi-wavelength and  multi-messenger searches for pulsar counterparts easier, as they may provide timing solutions valid for the epoch of other instruments' observations (including e.g. the long time spans covered by the LIGO and VIRGO detectors for gravitational waves; see e.g. \citealt{LigoPSR2017}).
Besides the advantages for timing purposes, the availability of information on both radio and gamma-ray pulsations allows for a comparison between emission geometries (and eventually emission mechanisms) in these two very different energy ranges \citep[e.g.][]{Venter2012, Petri2021}.

In this paper we describe a multi-telescope and multi-wavelength timing campaign to precisely characterise the pulsar discoveries made as part of the TRAPUM shallow L-band survey of \textit{Fermi}-LAT unassociated point sources \citep{clark+23}.
The paper is organised as follows. In Sect. \ref{s:radio}, we describe the multi-telescope radio timing campaign. In Sect. \ref{s:fermi} we report the gamma-ray data, searches, and discoveries. In Sect. \ref{s:radio-gamma}, we describe a new method to jointly time pulse times of arrival obtained from the radio observations and gamma-ray photons. In Sect. \ref{s:results}, we present the multi-wavelength timing results and a brief comparison of radio and gamma-ray properties. In Sect. \ref{s:poln}, our basic polarimetric results are shown. Finally, in Sect. \ref{s:summary}, we present our summary and conclusions. 

\section{Radio timing campaign}
  \label{s:radio}

Some information on the initial phases of the timing campaign on discoveries made during TRAPUM's L-band survey of unassociated \textit{Fermi}-LAT sources was already reported in our previous paper describing the survey \citep{clark+23}.  We  presented an initial coherent timing solution for two pulsars, and basic spin, astrometric, and orbital parameters for the remaining seven. In the current work, we describe the full timing campaign in more detail.

In order to minimise telescope usage at MeerKAT, where no open time for time-domain experiments was available until September 2023, a multi-telescope timing campaign was set up to follow up TRAPUM discoveries starting in March 2021. For the brightest pulsars, with an  S/N of the integrated pulse profile above $\sim 20$ in the 10 minute discovery observation at MeerKAT, we obtained time at the Murriyang 64-m telescope (Parkes) in Australia. There we exploited the 3 gigahertz (GHz) bandwidth of the Ultra-Wideband Low receiver (UWL, \citealt{Hobbs2020+UWL}). For pulsars with a declination above $\sim -35$ degrees, we also obtained time at the Effelsberg 100-m radio telescope (Germany) and the NRT/Nan\c{c}ay decimetric radio telescope (France). Depending on the spectral properties of each pulsar and on their exact sky location, this usually allowed us to get a good detection with one- to two-hour pointings.

Starting in November 2021, 53 hours were also granted to us to follow up TRAPUM \textit{Fermi} discoveries at MeerKAT. This time was used to get initial parameters for our new pulsars, in order to better assess the suitability of less sensitive telescopes for their subsequent follow-up, and to obtain full timing solutions for the fainter and/or tighter orbit ones, for which the smearing of the pulse period due to binary motion over one to two hours would prohibit the detection of pulsations in an acceleration search.

Details on the observational set-up at each telescope (central frequency, bandwidth, frequency and time resolution, backend, de-dispersion type) are reported in Table \ref{tab:radiotelescopes}. 
\begin{table*}
  \caption{Observational parameters for the timing observations.}
  \centering
  \label{tab:radiotelescopes}
  \begin{tabular}{lcccccccc}
    \hline
    Telescope  & Freq  & BW    & No. chan & T$_{\rm{samp}}$ & Npol    & Nbit & Backend & Dedisp \\
               & (GHz) & (MHz) &          & ($\upmu$s)        &         &      &         &        \\
    \hline
    MeerKAT    & 1.3 & 856  & 1024 & 9.6 & 4 & 8  & PTUSE  & Coh \\
               & 0.8 & 544  & 1024 & 7.5 & 4 & 8  & PTUSE  & Coh \\
    Parkes     & 2.4 & 3328 & 3328 & 64  & 1 & 8  & Medusa & Coh \\
    Nançay     & 1.5 & 512  & 1024 & 64  & 1 & 4  & Nuppi  & Inc \\
    Effelsberg & 1.5 & 250  & 1024 & 82  & 1 & 32 & PSRIX  & Inc \\
               & 1.5 & 250  & 512  & 51  & 1 & 8  & EDD    & Inc \\
    \hline
  \end{tabular}
  \tablefoot{Left to right the columns report: telescope used, central observing frequency, useable observing bandwidth, number of frequency channels, sampling time, number of polarisations (1=total intensity, 4=full Stokes), number of digitising bins, backend used and de-dispersion type (coherent or incoherent)}
\end{table*}
Whenever possible, observations were executed with a pseudo-logarithmic cadence (a few observations on day 1 and single observations on days 2, 5, 10, etc.), followed by monthly observations, in order to efficiently achieve timing coherence.

The basic steps to obtain a timing solution for each pulsar are briefly described below.
First, the position of the pulsar was refined through a triangulation method based on the comparison of the S/N of the integrated pulse profiles of different detections (obtained in multiple neighbouring coherent tied-array beams used in the survey's discovery observations), given a model of the beam point spread function. This method made use of the SeeKAT software described in \cite{Seekat} and allowed us to achieve positional accuracy typically better than 5 arcseconds. The improved position was used in all the follow-up timing observations and allowed us to promptly investigate multi-wavelength counterparts of our discoveries (e.g. \citealt{Dodge2024_J1910} and Phosrisom et al. in prep.). For short-period binary pulsars, we used this interferometric position to search for possible optical counterparts in the \textit{Gaia} DR3 catalogue \citep{Gaia,Gaia+DR3}, in which case the \textit{Gaia} astrometric solution could be used as a highly precise, independent prior for subsequent analyses.

Subsequently, for each observation, the most prominent radio frequency interference (RFI) signals were removed using the routine \texttt{rfifind} from the \texttt{PRESTO}\footnote{\url{https://github.com/scottransom/presto}} software package \citep{presto}. The same software was used to de-disperse the cleaned data (with \texttt{prepdata}) and either perform a blind acceleration search on the resulting file (through \texttt{accelsearch}, \citealt{rem02}), or, once a preliminary ephemeris was available, directly fold it (\texttt{prepfold}). 

In the case of a binary pulsar, the initial orbital solution was found, using the first few observations, by fitting an ellipse in the acceleration-period plane (following \citealt{fkl01}, especially useful in case of detections widely separated in time) and later 
by fitting a sinusoidal modulation to the barycentric spin periods (using \texttt{fit\_circular\_orbit.py} from \texttt{PRESTO}).

Multiple times of arrival (ToAs; typically from one to a dozen, depending on the S/N) were extracted from each observation using \texttt{PRESTO}'s \texttt{get\_TOAs.py} by cross-correlating a profile template with each folded observation. The template  initially used was the highest S/N profile for each pulsar. Once a coherent timing solution was achieved, multiple observations were added together in phase, resulting in profiles from which we obtained analytic, noiseless templates by fitting multiple Gaussian components.

The software \texttt{Tempo2}\footnote{\url{https://bitbucket.org/psrsoft/tempo2/src/master/}} \citep{tempo2} was used to obtain a coherent timing solution by minimising the root mean square (rms) of the ToAs  residuals in an iterative way, with increasingly more precise  ephemerides being used to refine both the template and the ToAs (two iterations were used in most cases). Wherever needed, constant time offsets (or `jumps') were also fitted, together with the pulsar parameters, to account for observatory-related time delays between the different telescopes, and/or observing systems.

In some cases, when the coverage of ToAs in time, and/or the orbital coverage, was too sparse, we achieved timing coherence using a new method. This algorithm utilises the software \texttt{Dracula}\footnote{\url{https://github.com/pfreire163/Dracula}} \citep{Dracula}, is implemented through \texttt{Tempo} \citep{tempo} and is able to determine the exact number of pulsar rotations also in the longer observational gaps. This was particularly useful for wider orbit, fainter pulsars.

In Table \ref{tab:timing} we list, for each new pulsar, the span of radio data used to obtain the timing solution, the number and total duration of the timing observations at each telescope included in the analysis, and the total number of ToAs obtained. 

\begin{table*}
\fontsize{9}{11}\selectfont
  \caption{Main parameters of the radio timing campaign.}
  \centering
  \label{tab:timing}
  \begin{tabular}{c @{\hspace{1 truecm}} c @{\hspace{1 truecm}} cccc @{\hspace{1 truecm}} cccc @{\hspace{1 truecm}} cccc}
    \hline
    PSR & Data span & \multicolumn{4}{c}{N. obs} & \multicolumn{4}{c}{tobs} & \multicolumn{4}{c}{N. ToA} \\
        & (MJD)     & \multicolumn{4}{c}{}       & \multicolumn{4}{c}{(h)}  & \multicolumn{4}{c}{}       \\
        \hline
        &           & MK & PKS & NRT & EFF & MK & PKS & NRT & EFF & MK & PKS & NRT & EFF  \\
    \hline
J1036$-$4353 & 59536 -- 60015 & 20 & 15 & -- & -- & 1.7 & 29.8 & --  &  --  & 99  & 81  & -- & -- \\
J1526$-$2744 & 59304 -- 59432 & 2 &  6 &  7 & -- &  0.3 &  9.3 & 6.3 &  --  & 16  & 28  & 27 & -- \\
J1623$-$6936 & 59250 -- 59683 & 23 & 14 & -- & -- & 2.2 & 14.6 & --  &  --  & 140 & 38  & -- & -- \\
J1709$-$0333 & 59304 -- 60084 & 20 &  4 & -- &  2 & 1.8 &  6.1 & --  &  3.4 & 111 & 32  & -- & 8  \\ 
J1757$-$6032 & 59250 -- 59725 & 29 & 12 & -- & -- & 4.1 & 13.9 & --  &  --  & 274 & 36  & -- & -- \\ 
J1803$-$6707 & 59197 -- 59624 &  2 & 19 & -- & -- & 1.4 & 11.0 & --  &  --  & 56  & 148 & -- & -- \\
J1823$-$3544 & 59250 -- 59684 & 19 &  9 & 10 & -- & 1.8 &  9.7 & 7.4 &  --  & 192 & 32  & 35 & -- \\ 
J1858$-$5422 & 59250 -- 59759 & 37 &  6 & -- & -- & 6.0 &  6.1 & --  &  --  & 265 & 10  & -- & -- \\ 
J1906$-$1754 & 59250 -- 60148 & 43 &  5 & -- & 20 & 4.0 &  6.1 & --  & 48.9 & 208 & 5   & -- & 72 \\ 
    \hline
  \end{tabular}
  \tablefoot{Left to right the columns report: Pulsar name, data span of the radio timing solution and, for each telescope (left to right: MeerKAT, Parkes, Nan\c{c}ay, Effelsberg), number, total duration of the observations used to obtain the timing solution and the number of ToAs. Observations with no detections are not included in the total counts. Discovery observations have been included, whenever it has been possible to use them for timing.}
\end{table*}
The time span of each data set (also including, whenever possible, the survey discovery observation) was determined by the number of observations needed to either get a coherent timing solution, including the measurement of the pulsar spin period first derivative or to obtain a good enough ephemeris,  which would enable the detection of gamma-ray pulsations, as described in the next section.

\section{Gamma-ray observations and pulsation searches}
\label{s:fermi}
For each newly discovered MSP, upon obtaining an initial phase-connected radio timing solution, we performed a search for gamma-ray pulsations in the \textit{Fermi}-LAT data, and used these pulsations, when detected, to improve the timing ephemerides (see Sect.~\ref{s:radio-gamma}).
For these analyses, we used \textit{Fermi}-LAT data according to the `Pass 8' \texttt{P8R3\_SOURCE\_V3} instrument response functions \citep{Pass8,Bruel2018+P305}, with the same energy-dependent photon selections that were used to produce the 4FGL-DR3 source catalogue \citep{4FGL-DR3}. Gamma-ray pulsation searching and timing relies on photon probability weights, derived from a spectral and spatial model of the gamma-ray sky, to disentangle fore- and background events from photons emitted by the target pulsar \citep{Bickel2008,Kerr2011,Bruel2019+Weights}. These weights were computed using the \texttt{gtsrcprob} routine, using the positions and spectra of sources from the 4FGL-DR3  \citep{4FGL-DR3}, as well as the \texttt{gll\_iem\_v07.fits} and \texttt{iso\_P8R3\_SOURCE\_V3\_v1.txt} Galactic and isotropic diffuse emission models. 

The distribution of photon weights, $w$, for a source can be used to estimate the detectability of gamma-ray pulsations from a newly discovered pulsar, prior to searching. The significance of gamma-ray pulsations is typically quantified by the $H_M$-test \citep{,Kerr2011}, which incoherently sums the weighted Fourier power up to $M = 20$ harmonics. The expected value for $H_M$ is proportional to $W^2 = \sum_i w_i^2$ \citep{Nieder2020+Methods}, with a prefactor that depends on the pulse profile shape, but which is typically of order unity. The $W^2$ quantity can also guide a minimum cutoff ($w_{\rm min}$) on the photon weights, used to remove the large number of low-weight photons that do not contribute meaningfully to the pulsation detection to speed up computations during searching or timing. We chose $w_{\rm min}$ values for which 99\%\ of the full $W^2$ was retained, typically corresponding to $w_{\rm min} \approx 0.01$--$0.03$.

For four pulsars, PSRs~J1623$-$6936, J1709$-$0333, J1757$-$6032, and J1858$-$5422, the radio ephemerides obtained in Sect.~\ref{s:radio} were sufficiently accurate that a single fold of the \textit{Fermi}-LAT data revealed pulsations reaching back to the beginning of the \textit{Fermi} mission in 2008, albeit with some phase drifts requiring small adjustments to the spin frequency, $f,$ and its first derivative, $\dot{f,}$ which were consistent with the radio-derived uncertainties on these parameters. For a fifth MSP, PSR~J1803$-$6707, whose gamma-ray pulsations were presented in \citet{clark+23}, significant pulsations were detected within the epochs covered by the radio timing solution but unpredictable orbital period variations prevented an extrapolation to earlier data. We address the gamma-ray timing for this pulsar in Sect.~\ref{s:j1803}.

For one of these pulsars, the isolated MSP J1709$-$0333, pulsations were initially detected only faintly (with $H=44$, after tweaking $\dot{f}$ slightly), and the resulting pulse profile contains a significant unpulsed component above the estimated background level, indicating that photon weights may be overestimated. In addition, the timing position for this pulsar lies 5 arcmin from the corresponding 4FGL-DR3 source, 4FGL J1709.3$-$0328, well outside the 4 arcmin 95\% localisation ellipse. 
Taken together, these features suggest that the pulsar and background gamma-ray spectra are mis-modelled in this region, with some background flux perhaps being misattributed to 4FGL J1709.3$-$0328 in the model. 
This is perhaps unsurprising as, at Galactic coordinates $(l,b)=(17.4\degr,+20.7\degr)$, this pulsar lies in a particularly complex region of the diffuse gamma-ray sky, along the edge of the northern \textit{Fermi} Bubble \citep{Su2010+Bubbles,Ackermann2014+Bubbles} and in a region with a significant dust excess \citep{Fermi_IEM}. 
To address this, we recomputed the photon weights with \texttt{gtsrcprob} using the pulsar timing position instead of the 4FGL position, and then applied an energy-dependent re-weighting to maximise the $H$-test, similar to the ``model weights'' method developed by \citet{Bruel2019+Weights}. 
For this, we used the photon re-weighting equation derived by \citet{Kerr2019}, $w^\prime = s w / (s w + 1 - w)$, multiplying the source flux with a power-law spectral function $s(E) = A (E / 1000\, \textrm{MeV})^\gamma$, and varied the overall scale factor $A$ and spectral index $\gamma$ to maximise the re-weighted $H$-test. 
We found that $A = 0.3$ and $\gamma = 0.7$ results in the maximum $H = 76$, indicating that the low energy spectrum of the pulsar/background is over/under-estimated, respectively, or some combination thereof, in the 4FGL-DR3 model. 

For the other MSPs the validity intervals of the radio solutions were too short for significant gamma-ray pulsations to be detectable within their validity period, and timing uncertainties were large enough that phase connection was not maintained when extrapolating to earlier gamma-ray data. In these cases it was necessary to perform a `brute force' search over the uncertain parameters ($f$, $\dot{f}$, $P_{\rm b}$, $\alpha$ and $\delta$). 
To save computing resources and given that most pulsars have the most Fourier power in the lowest harmonics \citep{Pletsch2014+Methods}, we chose to truncate $H_M$ at $M=3$. Significant detections have been claimed when $H_M > 25$ from a single trial \citep{Smith2019}, but much larger values are required to overcome the `trial factors' that arise when a range of parameters must be searched over \citep[e.g.][]{Nieder2019}. 

As already reported in \citet{clark+23}, PSR~J1526$-$2744 was detected in this way, but gamma-ray pulsations were not detected from the remaining three MSPs, PSRs~J1036$-$4353, J1823$-$3543, and J1906$-$1754, despite these efforts. For the first two of these pulsars, this is most likely explained by the low gamma-ray flux. For PSR~J1036$-$4353, in particular, the radio timing ephemeris covers around 500 days, and contains four orbital frequency derivatives describing short-timescale variations in the orbital period. However, $W^2 = 134$ for this pulsar, meaning that we would not expect to detect pulsations on timescales shorter than around 3 years, but the orbital phase variations mean that the timing model becomes uncertain on far shorter timescales, preventing a detection of gamma-ray pulsations from this pulsar. PSR~J1823$-$3543 is even fainter, with $W^2 = 67$, and we would require a timing model covering nearly all of the \textit{Fermi}-LAT data to detect it. We suspect that its signal is buried in the noise of the large number of trials that were required in our search. The non-detection of PSR~J1906$-$1754 is harder to explain (see Sect. \ref{ss:otherbins}), as with $W^2 = 111$ we would expect to detect pulsations even in a 5D search around the radio ephemeris.

\section{Joint radio and gamma-ray timing}
\label{s:radio-gamma}
Radio and gamma-ray data are very different in various aspects, which directly affects the timing accuracy. The gamma-ray data are sparse, sometimes with millions of rotations between two individual gamma-ray photons from the pulsar, but cover the full $15$ years between the launch of the \textit{Fermi} satellite and today. Such data are well suited to measure pulsar parameters for which the precision depends directly on the data time span, namely, precision improves more rapidly than with the observing span $t^{-1/2}$, such as the spin-frequency derivative, $\dot{f}$, the orbital period, $P_{\rm b}$, or proper motion. The radio data give precise pulse arrival times, but have gaps and usually (unless archival data are found at the pulsar position) only span the time between discovery and the most recent timing observation. Due to their higher S/N, the radio data typically measure parameters, such as the sky position or the binary projected semi-major axis, with a higher precision.

Gamma-ray pulsations have been detected from six of the pulsars presented here; thus, using both data sets should increase sensitivity in the timing analysis for these. For several of these pulsars, (e.g. PSR~J1623$-$6936), the typical iterative timing approach of alternating radio and gamma-ray timing analyses did not converge on a consistent sky position, because of the relatively short time span of the radio data. To fix this, we used a timing approach in which we fit the pulsar parameters utilising both data jointly. 

\subsection{Joint timing technique}
Our joint radio and gamma-ray timing analysis code\footnote{\url{https://fermi.gsfc.nasa.gov/ssc/data/analysis/user/jrag-timing.py}} uses the pulsar timing software \texttt{PINT}\footnote{\url{https://github.com/nanograv/PINT}} \citep{PINT}. With \texttt{PINT} the radio and gamma-ray data are folded individually for a given set of pulsar parameters. For these, a joint log-likelihood function, $\log \mathcal{L}_{\rm j}$ is constructed from the two respective standard timing statistics -- the radio $\chi^2_{\rm r}$ statistic and the gamma-ray template-based log-likelihood $\log \mathcal{L}_{\rm g}$ -- as
\begin{equation}
    \log \mathcal{L}_{\rm j} = \log \mathcal{L}_{\rm p} - 0.5 \chi^2_{\rm r} + \log \mathcal{L}_{\rm g} \,,
\end{equation}
where $\log \mathcal{L}_{\rm p}$ represents the log-priors. The radio $\chi^2_{\rm r}$ statistic is computed from the ToA residuals and uncertainties. $\log \mathcal{L}_{\rm g}$ is the commonly used `unbinned' log-likelihood statistic \citep{Kerr2015}. This parameter gives the likelihood that the distribution of pulsar phases at the emission times of the gamma-ray photons is described by a given template pulse profile, with the template consisting of wrapped Gaussian peaks \citep{Ray2011+Timing}.

We performed the full joint radio and gamma-ray timing analysis on four binary pulsars and one isolated pulsar. Using both radio and gamma-ray data, we fit for the astrometry, spin, and orbital parameters, as well as marginalising over the gamma-ray pulse-profile template \citep{Nieder2019}, using the Affine Invariant Markov chain Monte Carlo (MCMC) Ensemble sampler \texttt{emcee} \citep{emcee} to explore the parameter space. The timing solutions are presented in Tables~\ref{tab:ephemerisRB}, \ref{tab:ephemerisBIN}, and \ref{tab:ephemerisISO}, together with the radio-only solutions of the three pulsars with no gamma-ray pulsed counterpart.

\begin{figure*}
  \centering
        \includegraphics[width=0.7\textwidth]{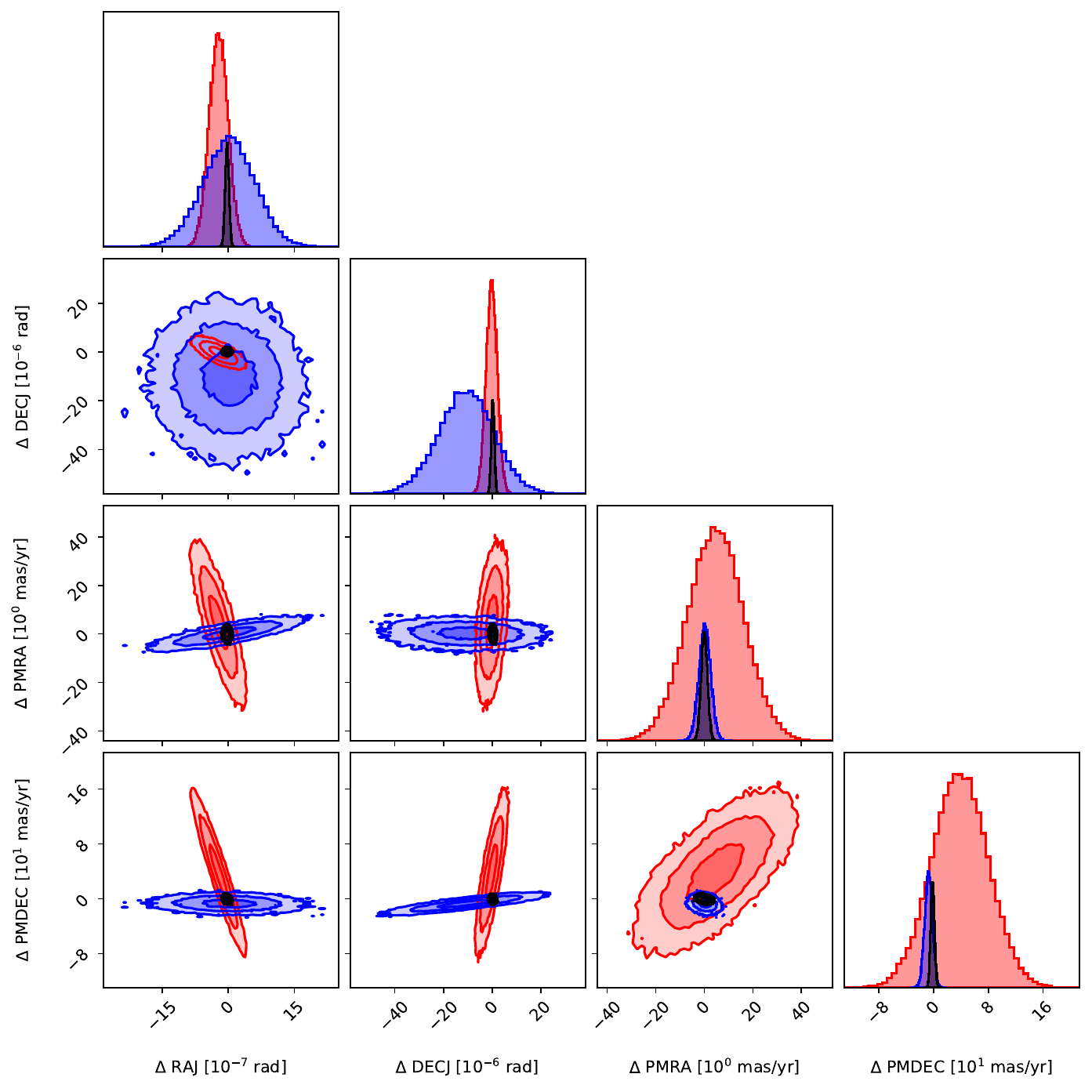}
        \caption{Corner plot for the astrometry parameters for PSR~J1858$-$5422, showing resulting posterior distributions for radio-only (red), gamma-ray-only (blue) and joint-radio-and-gamma-ray (black) timing analyses. The 1D posterior distributions on the diagonal are given in arbitrary units. The high S/N data allow us to pin down the positional parameters more precisely, while the proper motion parameters are measured more accurately by the long-term gamma-ray data. The sensitivity improvement of the joint analysis allows the precise measurement of all four parameters. The complete plot with posterior distributions for all pulsar parameters is shown in Fig.~\ref{f:j1858_compare}.}
    \label{f:j1858_astro}
\end{figure*}

Solving the problem of the non-convergence seen for some of the pulsars was not the only achievement of the joint radio and gamma-ray timing approach. The higher sensitivity enabled the measurement of the pulsars' proper motion, which was not significant in either data sets individually (illustrated in Fig.~\ref{f:j1858_astro}), making it possible to account for the kinematic contributions to the observed spin-down rate to more accurately estimate the energetics of these pulsars (see Sect.\ref{s:results}). The benefit of using radio and gamma-ray data jointly is demonstrated in the panels which show sky position parameters (RAJ, DECJ) versus the proper motion (PMRA, PMDEC). While the former are more accurately measured due to the higher S/N of the radio data, the proper motion measurement benefits from the longer time span of the gamma-ray data set.

\subsection{Gamma-ray timing of the redback PSR~J1803$-$6707}
\label{s:j1803}
Timing analyses of RBs are significantly complicated by long-term variations in their orbital periods, believed to be due to changes in the gravitational quadrupole moment of their companion stars \citep{Applegate1994+B1957,Lazaridis2011+J2051}. If they are not correctly accounted for, these lead to gamma-ray pulses apparently disappearing from data outside the range of validity of the radio ephemerides, which cannot be deterministically extrapolated over longer time spans. This effect led to us being initially unable to obtain a long-term gamma-ray timing solution for PSR~J1803$-$6707 beyond the interval covered by the radio timing in \citet{clark+23}. Since then, we have refined our methods for timing gamma-ray RBs, allowing us to resolve this. These methods, also used in \citet{Thongmeearkom2024+RBs}, build upon the procedure developed in \citet{Clark2021+J2039}, in which we treat the orbital phase variations as a stationary Gaussian process.

\begin{figure*}
  \centering
        \includegraphics[width=0.8\textwidth]{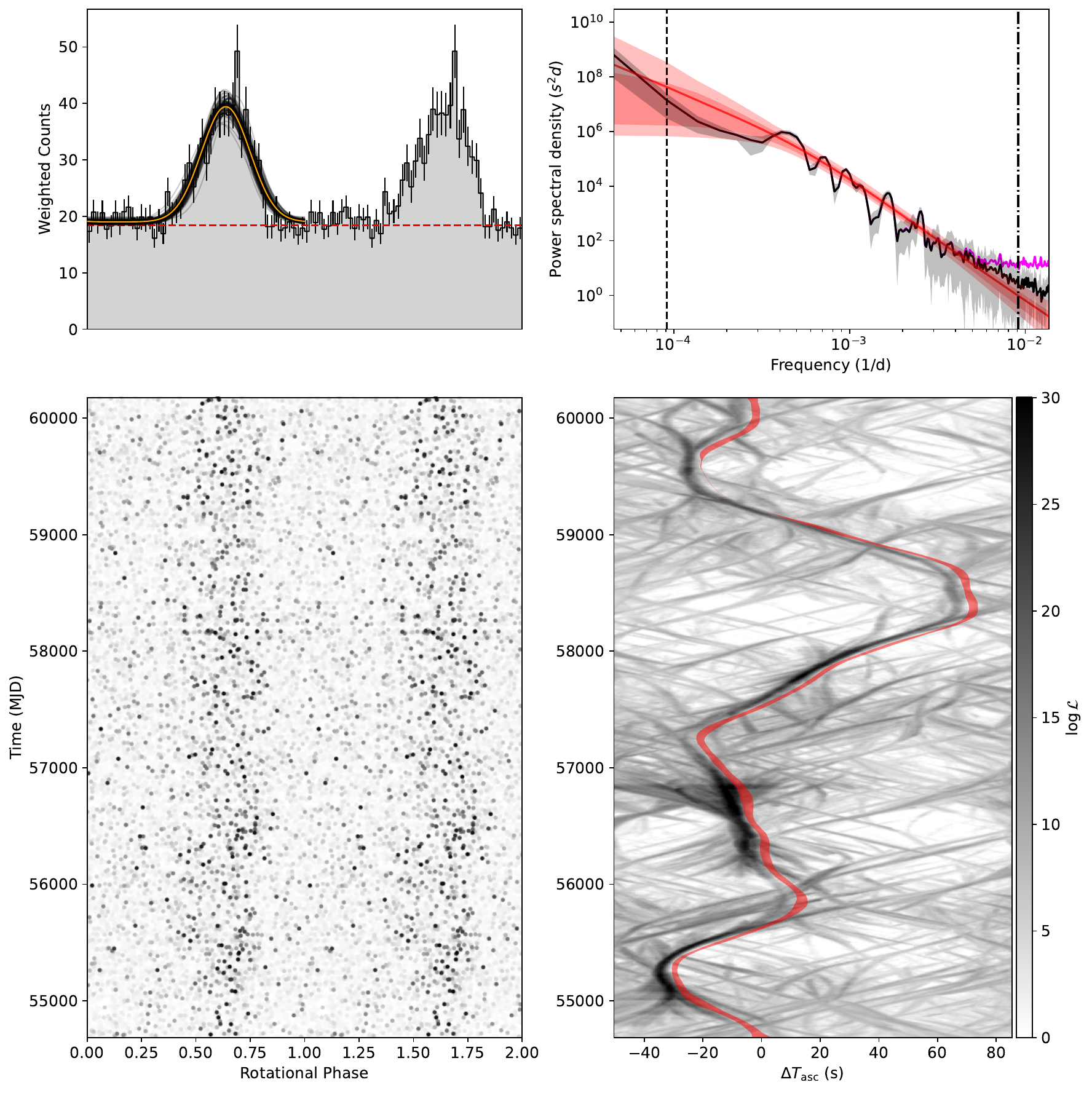}
        \caption{Gamma ray pulsations and orbital phase variations over the course of the \textit{Fermi}-LAT data for PSR~J1803$-$6707. The panels on the left show the weighted gamma-ray photon phases for the highest-likelihood timing solution (lower panel) and the integrated pulse profile (upper panel). The orange and the 100 faint black curves represent the associated highest-likelihood pulse-profile template and randomly drawn samples from the Monte Carlo analysis to visualise the uncertainty on the pulse-profile template. The bottom-right panel shows the orbital phase variations as a function of time. The grey-scale image shows the log-likelihood for offsets from the pulsar's $T_{\rm asc}$ as measured in overlapping $800$-day windows. The red curves represent the $95\%$ confidence interval on those deviations, obtained from Monte Carlo timing analysis, with a $5$\,s-offset for clarity. The top-right panel shows the power spectral density of the orbital phase variations. The red shaded regions illustrate the 68\% and 95\% confidence intervals of the Mat\'{e}rn model. The black curve and grey shaded region show the estimated power spectrum and its 95\% confidence interval from the joint radio and gamma-ray fitting, while the magenta curve, mostly hidden behind the black one, shows the power spectrum estimated from gamma-ray timing alone. The magenta curve is added to emphasise the increased sensitivity at higher frequencies caused by utilising the radio data.}
    \label{f:J1803_fermi_timing}
\end{figure*}

Starting from the radio ephemeris presented in \citet{clark+23}, we first searched for pulsations throughout the gamma-ray data, using a $500$-day sliding window to strike a balance between including enough data to detect significant pulsations, while also being robust against short-timescale variations in the orbital phase. This search spanned small ranges in epoch of ascending node, $T_{\rm asc}$, orbital period $P_{\rm b}$, and the first derivative of the orbital frequency, {\it FB1}. This resulted in a connected `track' of detections dating back to the start of the \textit{Fermi}-LAT data, to which we could fit an initial Gaussian process model describing the orbital phase changes. The results of this sliding window approach are illustrated in Fig.~\ref{f:J1803_fermi_timing}. We used this same sliding window approach to try to find gamma-ray pulsations from PSR~J1036$-$4353 but did not detect significant pulsations, which is unsurprising given the low weighted-photon count from this pulsar. 

Using this as starting point, we then performed a full gamma-ray timing analysis in which we jointly fit the parameters of the template pulse profile and the timing model (astrometry, spin and orbital parameters), alongside the orbital phase variation Gaussian process and three `hyperparameters' describing its covariance function. We adopted the commonly used Mat\'{e}rn covariance function, which is parameterised by an amplitude, $h$, length scale, $\ell$, and a smoothness parameter, $\nu$. This covariance function corresponds to a noise power spectral density that follows a smoothly broken power law, that is flat below a corner frequency $f_c = \sqrt{\nu}/\sqrt{2}\pi \ell$, breaking to a power-law at higher frequencies with a spectral index of $\alpha = -(2\nu + 1)$. As described in \citet{Thongmeearkom2024+RBs} we used a Gibbs sampling approach to deal with the multi-component likelihood function for each photon. This approach also allows radio ToAs to be incorporated in the sampling. Details on these methods and a public release of the code will be presented in future works (in preparation).

Since PSR~J1803$-$6707 has a bright optical counterpart \citep[see][]{clark+23}, we used \textit{Gaia} DR3 measurements as priors on the astrometric parameters (position, proper motion and parallax). In this case, the gamma-ray timing is not sensitive enough to improve upon these. The results of this timing procedure are shown in Fig.~\ref{f:J1803_fermi_timing}, and the full gamma-ray timing solution is provided as supplementary material on Zenodo (see Sect. \ref{sec:dataAvail}) in two versions ('ORBIFUNC' and 'ORBWAVES') that can be used with \texttt{Tempo2} and \texttt{PINT}, respectively\footnote{The ORBIFUNC parameters (for \texttt{Tempo2}) are a time-domain interpolating function describing the orbital phase variations over time, while the ORBWAVES parameters (for \texttt{PINT}) are the Fourier-domain representation of the same function.}.

\section{Timing results}
\label{s:results}

The results of the timing campaign are summarised in Tables \ref{tab:ephemerisRB}, \ref{tab:ephemerisBIN}, and \ref{tab:ephemerisISO}, where the best set of measured and derived timing parameters are reported for the two RBs, the other binary pulsars and the isolated MSP J1709$-$0333, respectively. Pulsars whose timing parameters have been obtained from the joint analysis of the radio ToAs and gamma-ray photons are in boldface font. Barycentric Dynamical Time (TDB) scale and the DE421 JPL Solar system ephemeris were used for the timing analysis of all pulsars.

\begin{table}
\fontsize{8}{10}\selectfont
  \centering
\caption{Timing parameters for the RBs J1036$-$4353 and J1803$-$6707.}
  \label{tab:ephemerisRB}
  \begin{tabular}{lll}
    \hline
    \multicolumn{3}{c}{Timing parameters}\\
    \hline
                                 & J1036$-$4353        & \textbf{J1803$-$6707}
       \\
    \hline
    $\alpha$ (h:m:s)                   & 10:36:30.215127(14) & 18:03:04.235339(10)
\\
    $\delta$ (deg:':'')               & $-$43:53:08.7252(3) & $-$67:07:36.15763(15)
\\
    $\mu_{\alpha}$  (mas yr$^{-1}$)                & $-$11.6(3)         
& $-$8.43(13)           \\
    $\mu_{\delta}$  (mas yr$^{-1}$)               & 2.9(3)              &
$-$6.3(1)           \\
    $\varpi$ (mas)                     & 0.4(4)              & 0.2(3)   
          \\
    {\it PosEpoch} (MJD)               & 57388               & 57388    
          \\
    $f$ (Hz)                      & 595.1998208151(4)   & 468.46771214902(10)
  \\
    $\dot{f}$ (Hz s$^{-1}$)                    & $-$2.26(3)$\times 10^{-15}$
     & $-$4.0434(6)$\times 10^{-15}$      \\
    {\it PEpoch} (MJD)                 & 59646.78438         & 59364.893677
        \\
    {DM} (pc cm$^{-3}$)               & 61.119(3)           & 38.382(3) 
         \\
    {\it Binary}                       & BTX                 & ELL1     
           \\
    $P_{\rm b}$ (d)                     & 0.259621257(6)      & 0.38047333(11)
     \\
    $x$ (lt-s)                    & 0.664789(3)         & 1.061914(2)   
    \\
    $T_{\rm{asc}}$ (MJD)                   & 59536.3052068(8)    & 59020.9969(5)
   \\
    {\it FB1} (Hz s$^{-1}$)                   & $-$2.3(3)$\times 10^{-18}$
      & - \\ 
    {\it FB2} (Hz s$^{-2}$)      & 7.9(6)$\times 10^{-25}$          & - \\

    {\it FB3} (Hz s$^{-3}$)      & $-$1.12(7)$\times 10^{-31}$      & - \\

    {\it FB4} (Hz s$^{-4}$)      & 6.0(4)$\times 10^{-39}$          & - \\

    {RMS} ($\mu$s)                 & 14.803              & 11.547       
      \\
    \hline
    \multicolumn{3}{c}{Hyperparameters of orbital-phase-covariance function}\\
    \hline
    $h$ (s)                      & -                                & $17^{+7}_{-10}$
\\
    $\ell$ (d)                   & -                                & $>
560$ \\ 
    $\nu$                        & -                                & $1.84^{+0.44}_{-0.41}$
\\
    \hline
    \multicolumn{3}{c}{Derived parameters}\\
    \hline
    $P$ (ms)                      & 1.6801080326781(12) & 2.1346188308481(12)
\\
    $\dot{P}_{\rm int}$ (s s$^{-1}$)         & 5.50$\times 10^{-21}$    
       & 1.84$\times 10^{-20}$            \\
    $\dot{E}$ (erg s$^{-1}$)            & 4.58$\times 10^{34}$          
  & 7.47$\times 10^{34}$             \\
    $B_{\rm S}$ (G)              & 9.73$\times 10^{7}$            & 2.01$\times
10^{8}$             \\
    {\it M}$_{2, \rm min}$ (${\rm M}_\odot$) & 0.23                & 0.29
               \\
    $d_{\rm{NE2001}}$ (kpc)           & 2.1                 & 1.2       
         \\
    $d_{\rm{YMW16}}$ (kpc)            & 0.4                 & 1.4       
         \\
    $\mu$ (mas yr$^{-1}$)                  & 12.0(3)             & 10.6(3)
            \\  
    $v_{\perp}$ (km s$^{-1}$)           & 119                 & 60      
           \\ 
        S$_{1300}$ (mJy)            & 0.14 & 0.17 \\ 
    \hline
  \end{tabular}
  \tablefoot{The top part of the table reports the measured timing parameters.
From top to bottom we list: right ascension ($\alpha$), declination ($\delta$),
proper motion in $\alpha$ and $\delta$, parallax ($\varpi$), epoch of the
reported position, spin frequency ($f$), spin frequency derivative, epoch
of the reported $f$, dispersion measure ({DM} taken from the best detection
obtained at 800 MHz at MeerKAT), binary model used \citep{tempo2}, orbital
period ($P_{\rm b}$), projected semi-major axis ($x$), time of the ascending
node ($T_{\rm asc}$), four orbital frequency derivatives and the \texttt{Tempo2}
rms of the timing residuals. Position, proper motion and parallax are taken
from \textit{Gaia} DR3 \citep{Gaia+DR3}. For each parameter, the (\textit{Gaia}
or timing) one-sigma error on the last quoted digit(s) is reported in parenthesis.
The hyperparameters of the orbital-phase-covariance function are the amplitude,
$h$, the length scale, $\ell$, and the Mat\'{e}rn degree, $\nu$. Their values
are reported with 95\% confidence intervals. The bottom part of the table
lists the derived parameters. Top to bottom: spin period, intrinsic spin
period derivative $\dot{P}_{\rm int}$ corrected for the Shklovskii and Galactic
potential effects, spin-down power and surface magnetic field (calculated
using the corrected $\dot{P}_{\rm int}$), minimum companion mass (assuming
a neutron star mass of 1.4 M$_\odot$), {DM}-derived distances (using the
NE2001 and YMW16 models), total proper motion and transverse velocity, using
the NE2001 {DM}-derived distance and estimate of the 1300-MHz flux density.
The full ephemeris for J1803$-$6707 is available as supplementary material
on Zenodo (see ref{sec:dataAvail}.}
\end{table}

\begin{table*}
\fontsize{7}{9}\selectfont
  \centering
  \caption{Measured and derived timing parameters for the non-eclipsing binary
pulsars.}  
  \label{tab:ephemerisBIN}
  \begin{tabular}{lllllll}
    \hline
    \multicolumn{7}{c}{Timing parameters}\\
    \hline
                             & \textbf{J1526$-$2744} & \textbf{J1623$-$6936}
& \textbf{J1757$-$6032} & J1823$-$3543 & \textbf{J1858$-$5422} & J1906$-$1754
       \\
    \hline
    $\alpha$  (h:m:s)            & 15:26:45.096(2)   & 16:23:51.3868(6) 
  & 17:57:45.4474(3)    & 18:23:42.9989(4)   & 18:58:07.7664(3)    & 19:06:14.7894(4)
   \\
    $\delta$ (deg:':'')          & $-$27:44:06.02(9) & $-$69:36:49.254(3)
& $-$60:32:12.300(4)    & $-$35:43:40.879(13)    & $-$54:22:15.527(74   
& $-$17:54:34.31(4)     \\
    $\mu_{\alpha}$ (mas yr$^{-1}$)      & $-$12(4)          & $-$5.2(11)
        & $-$2.3(14)               & --                 & $-$8.7(14)    
          & --                  \\
    $\mu_{\delta}$  (mas yr$^{-1}$)     & $-$4(14)          & $-$7.1(13)
         & $-$3.0(23)               & --                 & 5(3)         
      & --                  \\
    {\it PosEpoch} (MJD)         & 59355.468037      & 59332            
  & 59400               & 59091              & 59563.356205        & 59700
              \\    
    $f$ (Hz)                     & 401.7446020972(2) & 414.95962743561(13)
& 343.34786701089(10) & 421.3525986326(14)      & 424.53897607040(8)   &
347.66288115516(17) \\
    $\dot{f}$ (Hz s$^{-1}$)             & $-$5.724(10)$\times 10^{-16}$ 
& $-$1.5598(7)$\times 10^{-15}$       & $-$3.529(6)$\times 10^{-16}$    
 & $-$3.76(5)$\times 10^{-15}$       & $-$7.420(5)$\times 10^{-16}$     
& $-$4.73(11)$\times 10^{-16}$       \\
    {\it PEpoch} (MJD)           & 59355.468037      & 59332            
  & 59400               & 59091              & 59563.356205        & 59700
              \\
    {DM} (pc cm$^{-3}$)         & 30.953(3)         & 46.414(3)         
 & 62.918(3)           & 81.686(3)          & 30.819(3)           & 98.090(3)
             \\
    {\it Binary}                 & ELL1              & ELL1             
  & ELL1                & ELL1               &   ELL1              & BT 
                \\
    $P_{\rm b}$ (d)            & 0.2028108285(3)   & 11.01366617(2)   
  & 6.280139735(6)     & 144.568639(3)      & 2.581951179(5)     & 6.4910850(3)
       \\
    $x$ (lt-s)                   & 0.224080(9)       & 6.728198(4)      
  & 9.612624(2)         & 51.733306(3)       & 1.6889079(11)         & 1.346404(7)
        \\
    $T_{\rm{asc}}$ (MJD)         & 59303.2059794(10) & 59192.9162236(12)
    & 59183.4010969(6)   & 59091.325976(8)   & 59564.9829160(6)   & 59700.935268(5)
    \\
    {\it EPS1}                   & --                & 2.28(14)$\times 10^{-5}$
 & 1.1(6)$\times 10^{-6}$  & 3.991(7)$\times 10^{-5}$     &   --        
       & --                  \\
    {\it EPS2}                   & --                & 4.54(9)$\times 10^{-5}$
  & $-$1.3(5)$\times 10^{-6}$  & 2.6631(10)$\times 10^{-4}$      &   -- 
              & --                  \\
    {RMS} ($\mu$s)           & 47.180            & 34.153              &
24.597              & 16.975             & 21.312              & 32.570 
            \\
    \hline
    \multicolumn{7}{c}{Derived parameters}\\
    \hline
    $P$ (ms)                     & 2.4891435871939(19) & 2.4098729945848(8)
& 2.9124980699779(7)  & 2.373309202899(8) & 2.3554963298210(9) & 2.8763496312214(15)
\\
    $\dot{P}_{\rm int}$ (s s$^{-1}$)    & 3.88$\times 10^{-21}$         
  & 8.57$\times 10^{-21}$            & 2.24$\times 10^{-21}$           &
2.08$\times 10^{-20}$        & 3.70$\times 10^{-21}$        & 3.61$\times
10^{-21}$       \\
    $\dot{E}$ (erg s$^{-1}$)            & 6.48$\times 10^{32}$          
 & 2.42$\times 10^{34}$             & 3.58$\times 10^{33}$             &
6.15$\times 10^{34}$            & 1.12$\times 10^{34}$             & 5.98$\times
10^{33}$             \\
    $B_{\rm S}$ (G)              & 7.83$\times 10^{7}$           & 1.45$\times
10^{8}$             & 8.17$\times 10^{7}$            & 2.25$\times 10^{8}$
          & 9.44$\times 10^{7}$            & 1.03$\times 10^{8}$        
   \\
    {\it M}$_{2, \rm min}$ (${\rm M}_\odot$) & 0.08  & 0.19             
  & 0.42                & 0.26               & 0.12                & 0.05
               \\
    $e$                          & < 2.2$\times 10^{-4}$               &
5.00(13)$\times10^{-5}$           & 1.5(8)$\times10^{-6}$   & 2.6928(10)$\times
10^{-4}$      &   < 9.7$\times 10^{-6}$      & --                  \\
    $\omega$ (degrees)           & --                & 26.0(20)         
     & 44(25)                  & 171.477(16)          &   --            
   & --                  \\
    $d_{\rm NE2001}$ (kpc)       & 1.2               & 1.3              
  & 1.8                 & 2.1                & 1.0                 & 2.9
                \\
    $d_{\rm YMW16}$ (kpc)        & 1.3               & 1.3              
  & 3.5                 & 3.7                & 1.2                 & 6.8
                \\
    $\mu$ (mas yr$^{-1}$)               & 9(7)              & 8.6(18)   
         & 8(4)                & --                 & 9(3)              
 & -- \\ 
    $v_{\perp}$ (km s$^{-1}$)           & 51                & 53        
         & 65                  & --                 & 45                
 & -- \\ 
    S$_{1300}$ (mJy)            & 0.07 & 0.10 & 0.06 & 0.17 & 0.06 & 0.12
\\
    \hline
  \end{tabular}
  \tablefoot{For each parameter, the \texttt{Tempo2} one-sigma error on the
last quoted digit(s) is reported in parenthesis. The parameters listed (except
for the non measured parallax and orbital frequency derivatives) are the
same as in Table \ref{tab:ephemerisRB}. For two pulsars the ELL1 binary model
\citep{ELL1} was used to fit orbits with low eccentricities ($e$); the reported
values for $e$ and the longitude of periastron $\omega$ have been derived
from the {\it EPS1} and {\it EPS2} Laplace parameters given by the model,
the limits are 95\% upper limits. Pulsars whose names are in bold have been
jointly timed in radio and gamma-rays. The reported spin period derivative
(and derived parameters) have been corrected for the Shklovskii effect whenever
a significant total proper motion $\mu$ was measured.}
\end{table*}

\begin{table}
\fontsize{8}{10}\selectfont
  \centering
  \caption{Timing parameters for the isolated MSP J1709$-$0333. }
  \label{tab:ephemerisISO}
  \begin{tabular}{ll}
    \hline
    \multicolumn{2}{c}{Timing parameters}\\
    \hline
                     & \textbf{J1709$-$0333}   \\
    \hline
    $\alpha$ (h:m:s)       & 17:09:32.65211(12)     \\
    $\delta$ (deg:','')   & $-$03:33:19.250(5)       \\
    $f$ (Hz)          & 283.76747691122(6)     \\
    $\dot{f}$ (Hz s$^{-1}$)        & $-$1.510(5)$\times 10^{-16}$       
  \\
    {\it PEpoch} (MJD)     & 59694                  \\
    {DM} (pc cm$^{-3}$)   & 35.791(6)              \\
    {RMS} ($\mu$s)     & 25.851                 \\
    \hline
\multicolumn{2}{c}{Derived parameters}\\
\hline
    $P$ (ms)             &  3.5240120216913(8) \\
    $\dot{P}_{\rm int}$ (s s$^{-1}$) &  1.89$\times 10^{-21}$ \\
    $\dot{E}$ (erg s$^{-1}$)   &  1.70$\times 10^{33}$ \\
    $B_{\rm S}$ (G)     &  8.25$\times 10^{7}$ \\
    $d_{\rm NE2001}$ (kpc)  &  1.4 \\
    $d_{\rm YMW16}$ (kpc)   &  0.2 \\
    S$_{1300}$ (mJy)   & 0.05 \\
    \hline
  \end{tabular}
  \tablefoot{For each parameter, the \texttt{Tempo2} one-sigma error on the
last quoted digit(s) is reported in parenthesis. The parameters listed (except
for the missing proper motion, compatible with zero, at 1-sigma, and the
orbital ephemeris) are the same as in the previous table.}
\end{table}

For PSR~J1036$-$4353, the lack of a gamma-ray pulsation detection and the relatively short interval spanned by the radio data make the Gaussian-process treatment (described in Sect. \ref{s:j1803}) unnecessary. Instead, we include four orbital frequency derivatives in the timing model  to obtain flat timing residuals over the radio data span. 

When reporting parameters derived from the spin-down rate of the pulsar (e.g. the surface magnetic field B and the spin-down energy $\dot{E}$), one needs to take into account that the observed first derivative of the pulsar spin period $\dot{P}_{\rm obs}$ is contaminated by the Galactic acceleration of the pulsar relative to the Earth and by its proper motion. The intrinsic value $\dot{P}_{\rm int}$ can be written as:
\begin{equation}
     \dot{P}_{\rm{int}}  = \dot{P}_{\rm{obs}} - \dot{P}_{\rm{Gal}} - \dot{P}_{\rm{Shk}}, 
 \end{equation}
where $\dot{P}_{\rm{Gal}}$ is the contribution due to the Galactic gravitational acceleration and $\dot{P}_{\rm{Shk}}$ is the Shklovskii effect \citep{Shklovskii}, a purely kinematic contribution to the observed period derivative, which we could compute for those pulsars for which a significant measurement of the proper motion was available either from the \textit{Gaia} catalogue or as a result of our joint timing. The Galactic contribution $\dot{P}_{\rm{Gal}} = P\, \frac{a_{\rm Gal}}{c}$ (where $P$ is the spin period of the pulsar and $a_{\rm Gal}$ is the acceleration in the Galactic potential) is calculated using the model from \citet{McMillan2017}.
The Shklovskii effect is expressed as:
 \begin{equation}
     \dot{P}_{\rm{Shk}} = \frac{P\, \mu^2\,d}{c} = \frac{P\,v^2_\perp}{c\,d}
 ,\end{equation}
where $\mu$ is the total proper motion, $d$ the distance to the pulsar, and $v_\perp$ the transverse velocity.

The distance estimates reported in Tables \ref{tab:ephemerisRB}, \ref{tab:ephemerisBIN}, and \ref{tab:ephemerisISO} are derived from the dispersion measure ({DM}) of the pulsars and from two different models (NE2001 and YMW16) of the distribution of free electrons in the Galactic interstellar medium \citep{NE2001, YMW16}. For the two RBs we can compare the {DM}-derived distances (considering a 30\% error on their estimates) with the 2 sigma lower limit derived from the \textit{Gaia} parallax measurement for the companion stars. For J1036$-$4353, the \textit{Gaia} distance ($> 0.9$ kpc) is incompatible with the YMW16 value ($0.4$\,kpc), while the lower limit for J1803$-$6707 (1.25 kpc) is in line with the {DM}-derived values from both models ($1.2$\,kpc and $1.4$\,kpc for NE2001 and YMW16, respectively). For the calculation of the transverse velocity $v_{\perp}$ and of $\dot{P}_{\rm{int}}$ we hence chose to use the distance derived from the NE2001 model in all cases.

\begin{figure}
  \centering
        \includegraphics[width=\columnwidth]{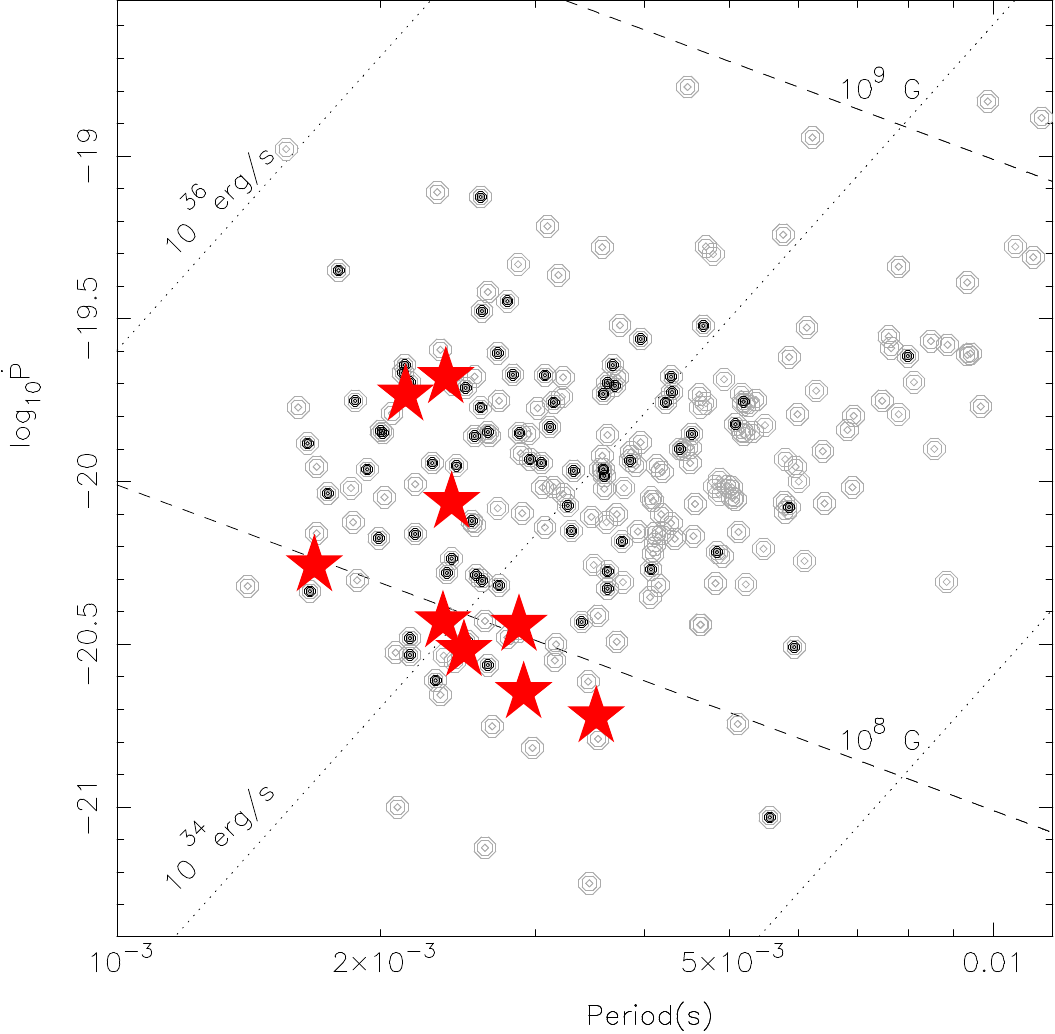}
        \caption{Period-period derivative diagram for Galactic millisecond
pulsars. Grey circles show  pulsars from PSRCat 1.70, while smaller black
points denote MSPs found targeting \textit{Fermi} unassociated point sources.
Red stars are the MSPs from this paper. The $\dot{P}$ used is, whenever possible,
the one corrected for the Shklovskii effect. Equal spin-down energy lines
are drawn as dotted and equal surface magnetic field lines are dashed.}
    \label{f:ppdot}
\end{figure}

Since timing data were not flux calibrated, an estimate of the flux density has been obtained using the radiometer equation modified for pulsars (see e.g. \citealt{RadiomEqPulsar}).
Figure \ref{f:ppdot} shows our nine MSPs as red stars in the period-period derivative diagram. As already noted in \cite{Ray2012+PSC} for the entire MSP population discovered in Fermi sources (drawn as black points in the diagram), the pulsars in our small sample appear to have shorter spin periods with respect to the general population of Galactic MSPs (plotted as grey circles and defined, here, as pulsars with spin period smaller than 10 ms). Our pulsars  have a median $P$ of 2.4 ms, 1.2 ms shorter than that of the Galactic population (and 0.4 ms smaller than that of the MSPs previously discovered targeting gamma-ray point sources). 

In the following, we highlight some interesting features of our timed targets.

\subsection{PSR~J1757$-$6032 and the detection of its Shapiro delay}

\begin{figure}
  \centering
        \includegraphics[width=\columnwidth]{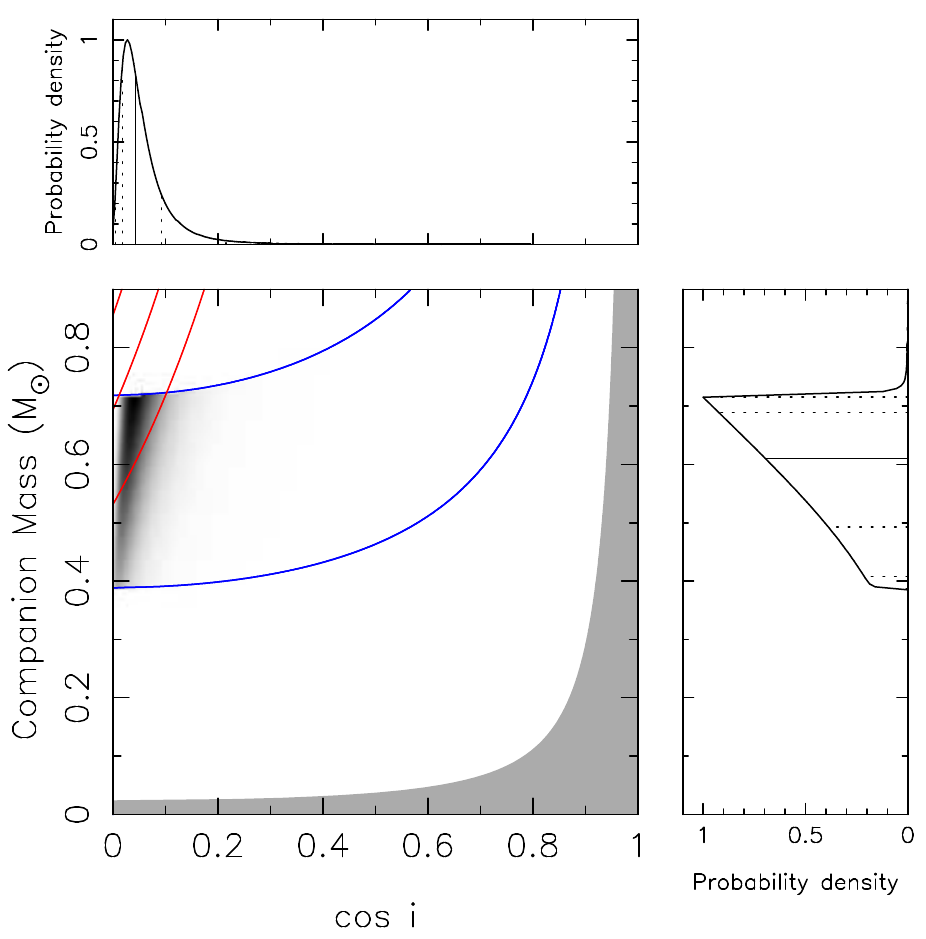}
       \caption{Companion mass and inclination constraints for the PSR~J1757$-$6032 binary system. The main panel shows the $M_{\rm 2}$-$\cos i$ plane, (with $M_{\rm 2}$ the mass of the companion star and $i$ the orbital inclination). The grey zone is excluded by the requirement that the pulsar mass must be positive. The gray scale shows the joint posterior probability density function (pdf), with white representing low or no probability. This pdf is limited, at the lower and upper sides, by pulsar masses of 1.17 and $3.2 \, \rm M_{\odot}$ (see text for justification); these are indicated by the blue lines. The red lines show the constraints derived from the median and $\pm 1$-$\sigma$ measurements of $h_3$, measured assuming $i = 87.5\, \deg$ (see text). The side panels include marginalized posterior probability distribution functions for $\cos i$ (top) and (to the right) $M_{\rm 2}$. The solid vertical lines indicate the medians of the distributions, the dotted lines indicate the 1-, 2-, and 3-$\sigma$ equivalent percentiles around the medians. Despite the lack of precise mass measurements, we can see that nearly edge-on orbital inclinations (i.e. with $\cos i$ near zero) are strongly preferred for the whole range of pulsar masses.}
    \label{f:J1757shap}
\end{figure}

This pulsar is in a binary system with an orbital period of 6.28 d with a very small orbital eccentricity and a relatively large projected semi-major axis of 9.6 light seconds. If we assume a pulsar mass of $1.4 \, \rm M_{\odot}$, then the minimum companion mass is $0.42 \, \rm M_{\odot}$. This is significantly larger than the prediction of \cite{1999A&A...350..928T} for the mass of a helium white dwarf (He WD) in a 6.3-d orbit around a MSP, which is about $0.24 \, \rm M_{\odot}$; this implies that the companion is likely a CO white dwarf. 

The basic characteristics of the system resemble PSR~J1614$-$2230 \citep{Demorest2010+J1614} and a few other systems like PSR~J1933$-$6211 (see detailed discussions in \citealt{2023A&A...674A.169G} and \citealt{2023MNRAS.520.1789S}). Compared to the other known pulsars with similarly high companion masses, these systems have 1 to 2 orders of magnitude faster spin periods, lower eccentricities (three orders of magnitude, on average) and smaller magnetic fields (up to three orders of magnitude).

These characteristics can be explained if the system originated in Case A Roche lobe overflow (RLO), namely, the pulsar started accreting mass from its companion when the latter was still in the main sequence \citep{Tauris2011+J1614}. In this model the unusually long accretion episode results in the fast spin, small B-fields and unusually small orbital eccentricities, even compared to those of MSP-He WD systems.

One of the reasons why these systems are interesting is that the measurement of the Shapiro delay \citep{1964PhRvL..13..789S} in PSR~J1614$-$2230 showed that the pulsar is quite massive \citep{Demorest2010+J1614},
a measurement that is important for the study of the EoS of nuclear matter at densities above that of the atomic nucleus (see e.g. \citealt{2016ARA&A..54..401O}). The latest mass measurements for this system (e.g. \citealt{2023MNRAS.520.1789S,ng15}) confirm a pulsar mass around 1.94 $\rm M_{\odot}$.

When studying the evolution of the system, \cite{Tauris2011+J1614} suggested that, despite its length, the accretion episode is not responsible for the high mass of PSR~J1614$-$2230, the latter is more likely a birth property. This was to some extent confirmed recently by the measurement of the mass of PSR~J1933$-$6211: despite its strong evolutionary similarities with PSR~J1614$-$2230 it has a mass around $1.4 \, \rm M_{\odot}$ \citep{2023A&A...674A.169G}. 
However, the precision of the mass of PSR~J1933$-$6211 still needs to be improved, and other similar systems, like PSR~J1101$-$6424, do not yet provide precise measurements \citep{2023MNRAS.520.1789S}. In order to start answering the question of whether neutron star (NS) masses are determined from their birth or by subsequent evolution, we need to measure more pulsar masses from this type (Case A RLO) of system. These are especially promising for detecting the role of accretion in the mass of the pulsar, not only because in this case the accretion episode is exceptionally long, but also because the relatively large companion masses and short spin periods (with resulting potentially high timing precision) make the measurement of the Shapiro delay easier. Furthermore, if accretion adds a significant amount of mass to the pulsar, we might find additional massive pulsars in this class of binaries, which might introduce additional constraints on the EoS.

Interestingly, it might be possible to make a mass measurement for PSR~J1757$-$6032. We have a significant detection of the orthometric amplitude of the Shapiro delay, $h_3 = 3.4 \pm 0.8 \, \rm \mu s$ (estimated assuming an inclination of 87.5 degrees, see below), however, fitting for both parameters of the Shapiro delay ($h_3$, $\varsigma$, \citealt{fw10}) yields no significant detection at present, which precludes precise constraints on the masses. The lack of precision of the Shapiro delay is caused by the insufficient precision of the radio timing (this is significantly worse for the gamma-ray timing) and the currently limited number of observations near superior conjunction.

To better quantify the detection, we made a quality-of-fit ($\chi^2$) map of the $M_{\rm 2}$-$\cos i$ plane (mass of the companion vs cosine of the orbital inclination), restricting the pulsar masses to the causality limit at the upper end ($3.2 \, \rm M_{\odot}$) and a value of $1.17 \, \rm M_{\odot}$ at the lower end, the lowest measured NS mass \citep{2015ApJ...812..143M}. We chose $\cos i$ because this quantity has a uniform probability for randomly aligned orbits, constituting a reasonable prior. From this $\chi^2$ map, we then derived a Bayesian joint posterior probability density function (pdf) and marginalised pdfs using the procedure described by 
\cite{2002ApJ...581..509S}. The results are displayed in Fig.~\ref{f:J1757shap}.
The main conclusion to take from this figure is that, for the whole range of NS masses we considered, the orbital inclination is always high, with a median value of $87.5^{+1.4}_{-2.8} \, \deg$ (68.3\% confidence level) and  $87.5^{+2.2}_{-10.0} \, \deg$ (95.4 \% confidence level). The range of companion masses (from 0.39 to about $0.75 \, \rm M_{\odot}$) results from this fact and the limits we set on the range of NS masses.

This high inclination indicates good prospects for the measurement of the masses of the components of this system in the near future via the two Shapiro delay parameters, provided that a) the timing precision improves and that b) a long MeerKAT observation can be made near superior conjunction as part of a dense orbital campaign.

\subsection{Redback pulsars J1036$-$4353 and J1803$-$6707} \label{ss:j1036+j1803}

\begin{figure}
  \centering
        \includegraphics[width=\columnwidth]{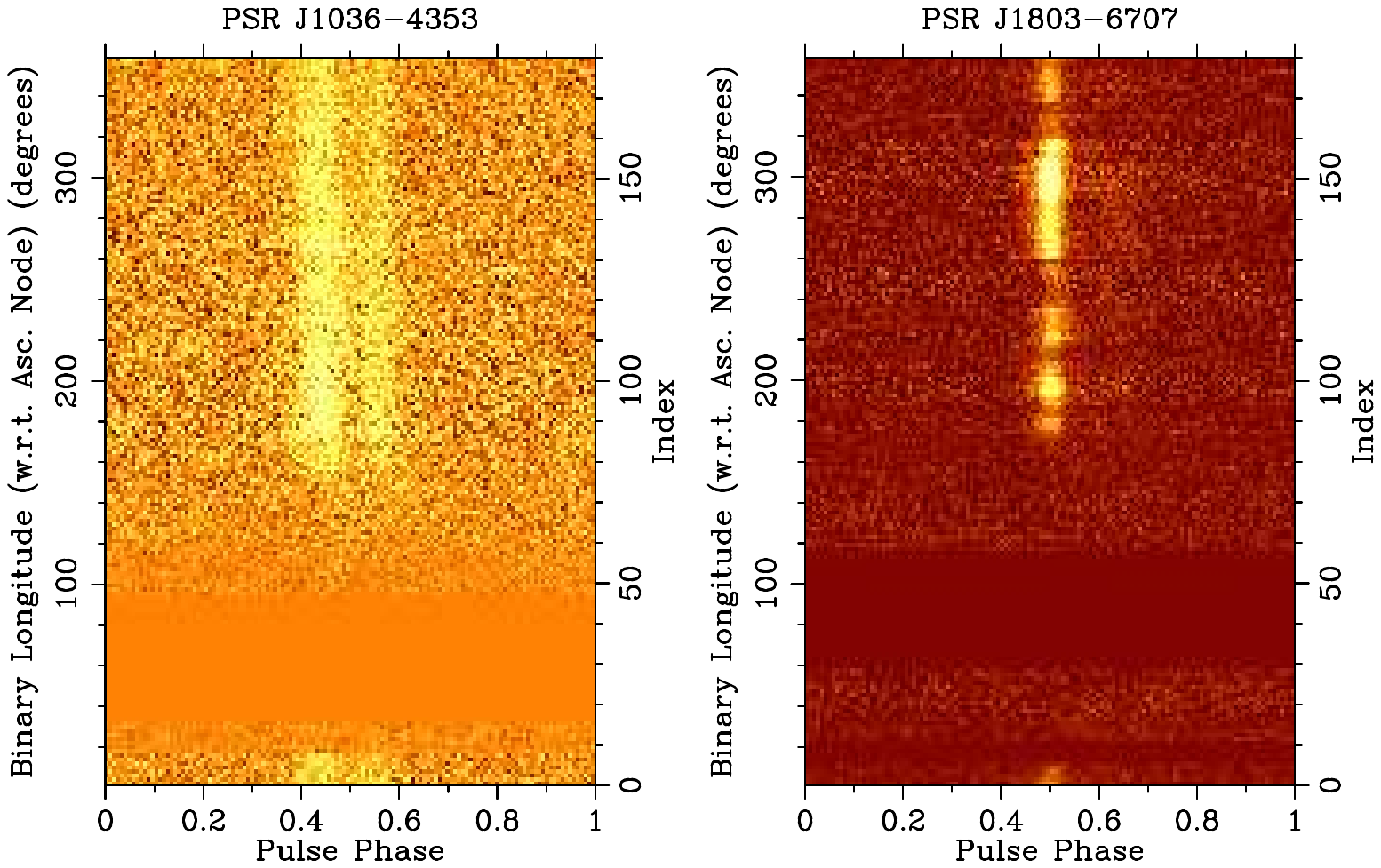}
        \caption{Waterfall plots showing, for the two RB pulsars J1036$-$4353 and J1803$-$6707, orbital longitude (with respect to the ascending node) versus pulse phase. The brighter the colour-scale, the stronger the signal. Orbital longitude ranges where the plots show a solid colour have no associated data, as observations were planned to avoid the eclipses. Data are from the Parkes telescope.}
    \label{f:RBeclipses}
\end{figure}

Figure \ref{f:RBeclipses} shows, for the two RBs J1036$-$4353 and J1803$-$6707, the orbital light curve obtained by summing all of the Parkes data. A very extended eclipse is visible between orbital longitudes $\sim 20\deg$ and $160\deg$ (orbital phases 0.05 to 0.45) for J1036$-$4353 and between longitudes $\sim 10\deg$ and $170\deg$ for J1803$-$6707. 

\begin{figure}
  \centering
        \includegraphics[width=\columnwidth]{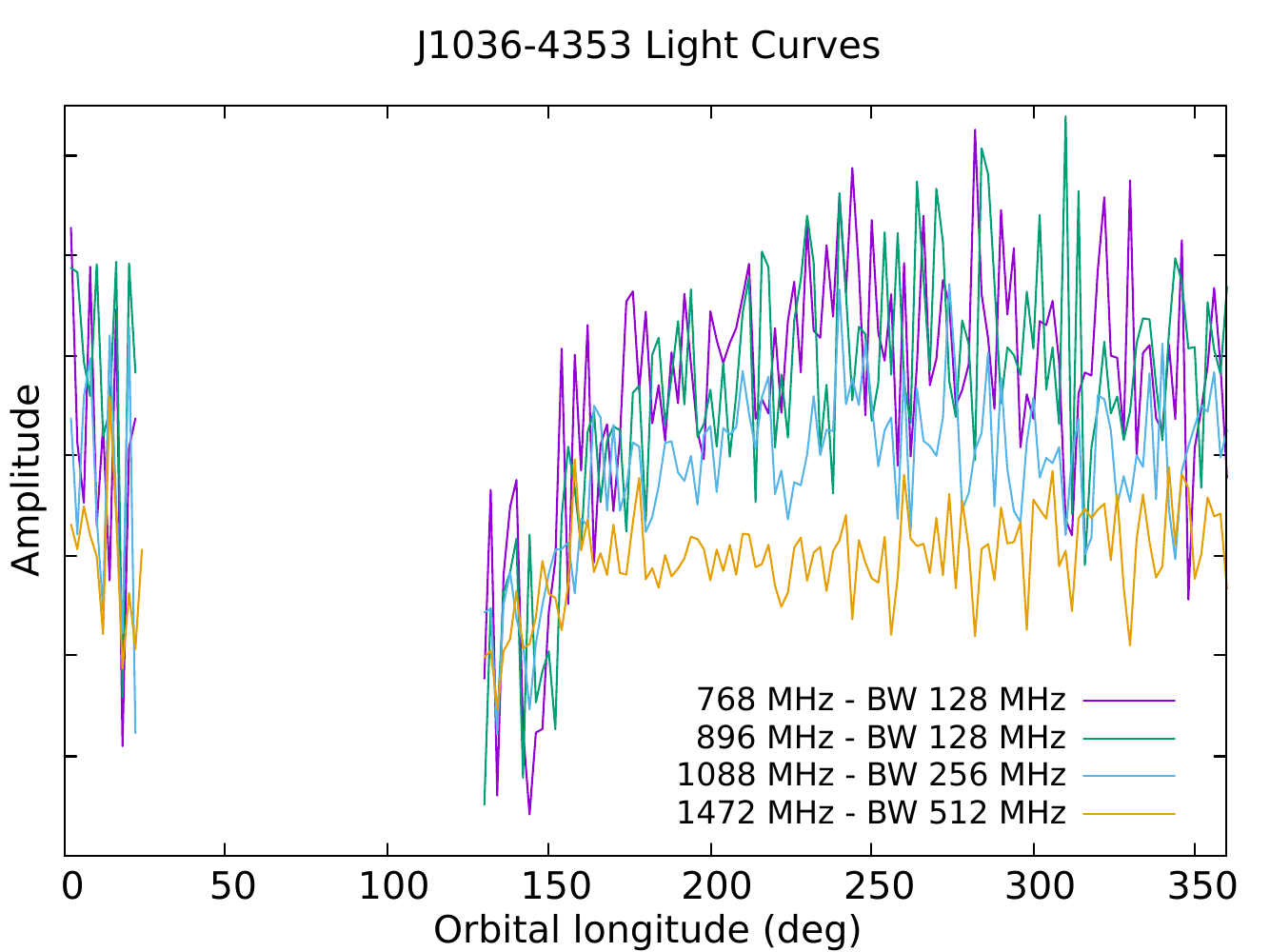}
        \caption{Amplitude of the pulsed radio emission as a function of the longitude from the ascending node for PSR J1036$-$4353 in four different frequency sub-bands. The amplitude is measured over a pulse phase range of  0.2 around  the peak. The central frequency and bandwidth of each sub-band are reported in the bottom-right corner. Points between longitudes 22$^{\circ}$ and 130$^{\circ}$ have not been plotted either because no data at all (between 34$^{\circ}$ and 94$^{\circ}$) or not enough data had been recorded, resulting in a much higher level of noise.}
    \label{f:J1036ecli}
\end{figure}

Thanks to the UWL receiver at Parkes, the availability of data over a 3.3 GHz-wide frequency band can in principle allow for a phase-dependent study of the eclipse duration; this would provide an interesting insight to supplement our understanding of the nature and geometry of the eclipsing material (see e.g. \citealt{polzin+20} and \citealt{Kansabanik2021}). 
For PSR J1036$-$4353, however, the S/N of the summed data is not sufficient to create bright enough phase-binned orbital light curves to reliably study these effects over more than $\sim 1$ GHz in bandwidth. Figure \ref{f:J1036ecli} shows the orbital light curve of the pulsed flux, obtained by selecting $\pm$ 0.1 in rotational phase around the peak, for the bottom part of the band (from 704 to 1728 MHz) split into four sub-bands with increasing bandwidth (to account for the steepness of the pulsar spectrum). No clear evidence for frequency evolution is visible. 

\begin{figure*}
  \centering
        \includegraphics[width=\textwidth]{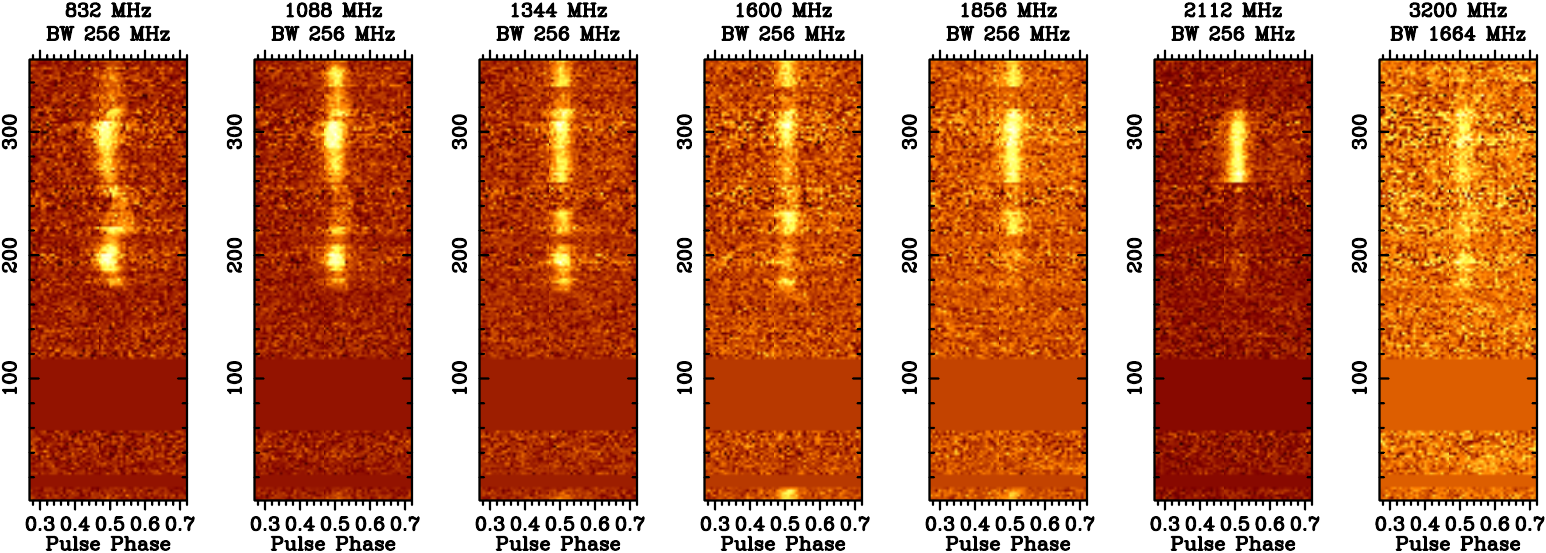}
        \caption{Waterfall plots representing the orbital light curves (longitude with respect to the ascending node vs pulse phase) for PSR~J1803$-$6707 in different frequency sub-bands. The central frequency and bandwidth are marked on the top of each panel. Only a pulse phase range of 0.4  around the peak is plotted.}
    \label{f:J1803ecli}
\end{figure*}

For J1803$-$6707, for which the S/N is high enough to produce orbital light curves in different sub-bands across the whole UWL band, the combination of a non-uniform orbital coverage and of strong interstellar scintillation, prevents us from measuring the frequency dependence of the eclipse duration. 
What may appear, in Fig.~\ref{f:J1803ecli}, as a frequency-dependent longitude of eclipse ingress, for instance (where the eclipse seems to begin later at intermediate frequencies), is actually due to the fact that the orbital longitudes below $\sim20\deg$ were observed only once (on MJD 59243). On that occasion, scintillation made the signal prominent almost exclusively at frequencies around 1.7 GHz (4th panel).
The same kind of effect is clearly visible in the 6th panel of Fig.~\ref{f:J1803ecli}, displaying 256\,MHz of band centred around 2.1\,GHz: here we see the effect of a very bright scintle present, at around 2\,GHz, in one (of only three) observations covering the orbital longitudes between 270\,$\deg$ and 320\,$\deg$). The phase variations visible in the lowest frequency sub-band plotted in the leftmost box of Fig.~\ref{f:J1803ecli} are likely the effect of dispersion measure delays caused by matter surrounding the system also away from the eclipse region; scintillation and profile evolution effects, however, can also contribute to create apparent pulse shape variations across the orbit, given the non-uniform orbital coverage.

\subsection{The other binary pulsars}
\label{ss:otherbins}

As already noted by \cite{clark+23} in the discovery paper, given its minimum companion mass and orbital period, PSR~J1526$-$2744 could also be a ``spider'' pulsar: either a relatively heavy BW, or a high-inclination RB (a large, albeit statistically unlikely, inclination would be required to both explain the small minimum companion mass with respect to the bulk of the RB population and the lack of visible eclipses in the system). The lack of a bright optical counterpart, also reported in \cite{clark+23}, seems, however, to favour the hypothesis that the companion star is a light (or on a highly inclined-orbit) He WD. 

Despite having a minimum companion mass also in the range typical of BWs, PSR~J1906$-$1754 has a much larger orbital period of 6.49 d and shows no eclipses. If the companion is a He WD, then the \cite{1999A&A...350..928T} relation predicts a mass of around $0.24 \, \rm M_{\odot}$.
If we assume a pulsar mass of $1.35 \, \rm M_{\odot}$, this results in an orbital inclination of $\sim13\deg$, which is a priori unlikely but should occur occasionally among MSP-He WD systems. Given the alignment of the spin with the orbital angular momentum one should expect for MSP-WD binaries, this would imply we are observing the pulsar from close to its spin axis. Two predictions stem from this: the first is that the radio pulse should be very broad, which is indeed observed: in the plot in Sect \ref{s:poln}, we see that this object has by far the widest pulse profile among all discoveries, with observable emission between spin phases of 0.2 to 1.0. Another consequence would be fainter observable gamma-ray emission, which is expected to be preferentially emitted around the pulsar's spin equator \citep[e.g.][]{Kalapotharakos2023}, as well as a weaker modulation of the gamma-ray profile; this could be the reason why no gamma-ray pulsations have been detected for this system.

This speculative picture would be strengthened if the observed gamma-ray flux suggested a particularly low efficiency for this system, but this is not the case here. Given the rather uncertain DM-model distances for this pulsar ($d_{\rm YMW16} = 6.8$\,kpc vs. $d_{\rm NE2001} = 2.9$\,kpc), the energy flux above 100MeV 4FGL~J1906.4$-$1757 ($G_{\rm 100} = 3.4 \times 10^{-12}$\,erg\,s$^{-1}$, \citealt{4FGL-DR4}) implies gamma-ray efficiencies ($\eta = 4\pi G_{\rm 100} d^2 / \dot{E}$) of 300\% (YMW16) or 60\% (NE2001). While apparent efficiencies greater than 100\% are possible if emission is strongly beamed towards us, the YMW16 value is far higher than any seen in a gamma-ray MSP \citep{3PC}, while the NE2001 value is more plausible, but still on the higher end of the population.  

Alternatively, the non-detection of gamma-ray pulsations from this pulsar, and also from PSR~J1823$-$3543, could be due to the pulsar not actually being the counterpart of the associated gamma-ray source. \citet{Kerby2023} searched for X-ray counterparts to our pulsars in \textit{Swift}-XRT data, finding, for 4FGL~J1906.4$-$1757 and 4FGL~J1823.8$-$3544, alternative X-ray sources whose positions are inconsistent with our pulsar timing positions, but are still within the \textit{Fermi}-LAT localisation regions. For 4FGL~J1906.4$-$1757 this X-ray source was deemed likely to be a blazar, which could provide an alternative explanation for the gamma-ray source. However, this picture requires the serendipitous discovery of two unrelated MSPs within the targeted 4FGL gamma-ray localisation region, a scenario which is rather unlikely. From the GBNCC survey discoveries \citep{McEwen2020}, \citet{Bhattacharyya2022} estimated a density of 0.002 MSPs per square degree that could be discovered by the GBNCC survey. This value suggests the chance of us finding a single unrelated MSP among the 79 \textit{Fermi}-LAT sources (with typical semi-major axes below $5\arcmin$) included in our survey was just 0.3\%, although this number could be somewhat higher given the relative sensitivity of our searches. 

Three pulsars, J1623$-$6939, J1823$-$3543, and J1858$-$5422, appear to have typical He WD companions, along with about half of the binary MSPs in the Galactic field. The first two have longer orbital periods (11 and 144 d, respectively) and a measurable eccentricity, while for the third, with an orbital period of 2.6 d, only an upper limit on the eccentricity ($e < 6 \times 10^{-6}$ ) is measurable. This is in line with the eccentricity-orbital period relation of \cite{pk94} valid for NS-WD binary systems formed via stable mass transfer through Roche lobe overflow. 

\subsection{Radio-gamma ray phase alignment}
\label{ss:cfr}

The relative phase offset between radio and gamma-ray pulses encodes information about a pulsar's emission and viewing geometry \citep[e.g.][]{Johnson2014,3PC}. For binary MSPs, assumed to have been spun-up via accretion from the companion star, and therefore spin-aligned to the orbit, this can also constrain the binary inclination angle \citep[e.g.][]{Corongiu2021+J2039}. Gamma-ray pulsar emission models predict a dependency between the phase lag ($\delta_{{\rm g}-{\rm r}}$) between the radio and gamma-ray pulses and the separation ($\Delta$) between the two main peaks in the gamma-ray pulse profile, with the largest $\Delta$ values expected to occur for pulsars whose radio and gamma-ray peaks are closely aligned with each other. Figure~10 in \citetalias{3PC} demonstrates that this correlation is significantly detected in the known population of gamma-ray pulsars.

\begin{figure*}
  \centering
        \includegraphics[width=\textwidth]{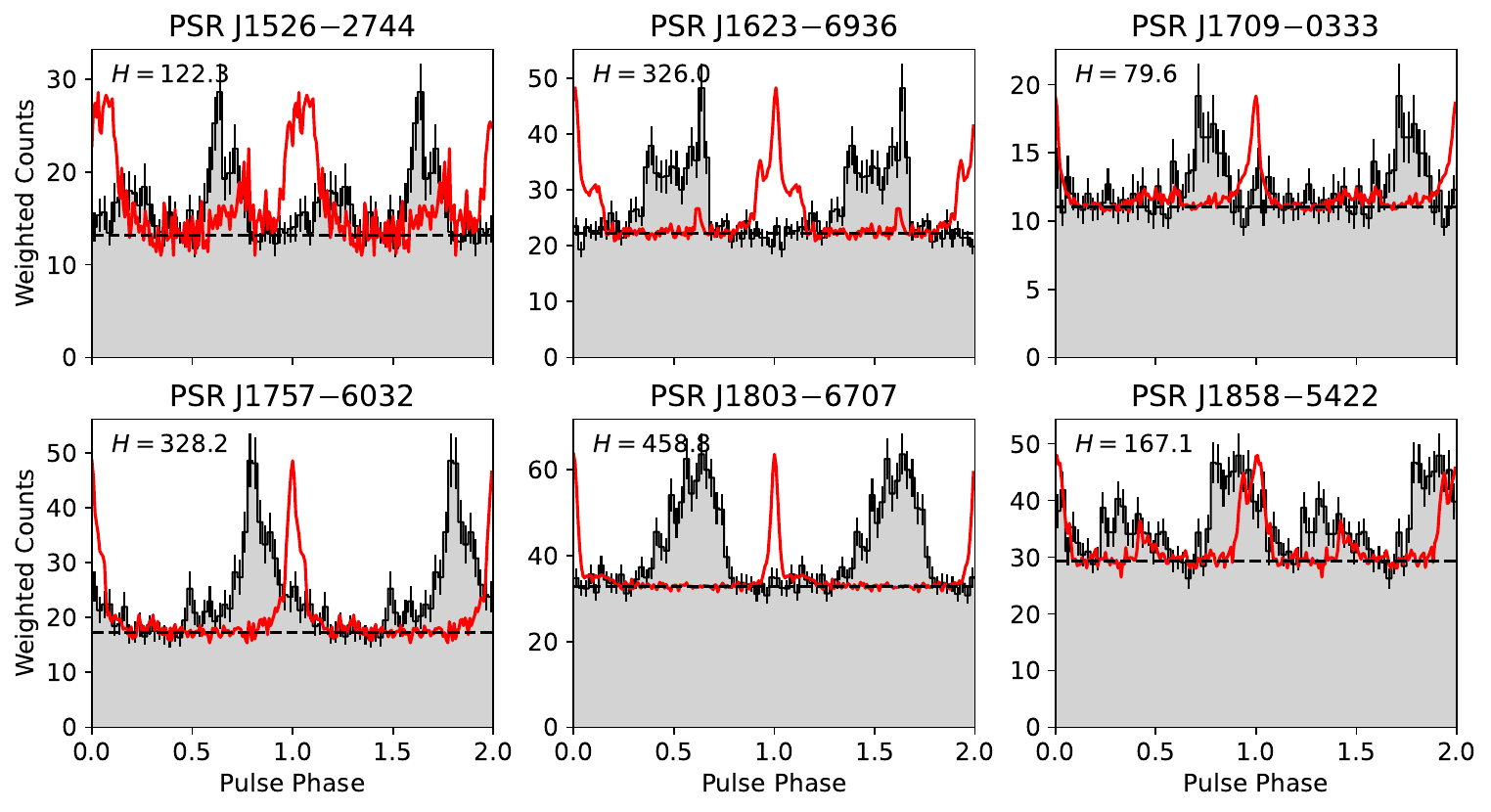}
        \caption{Phase-aligned radio (red curves) and gamma-ray (grey histograms) pulse profiles. The gamma-ray background level, estimated from the distribution of photon weights as $b = \sum_i w_i (1 - w_i) / n_{\rm bins}$, is shown by dashed horizontal black lines. Radio pulse profiles are in arbitrary flux density units, scaled to match the same pulse amplitude as the gamma-ray peak, and with the baseline flux matching the gamma-ray background level.}
    \label{f:phase_alignment}
\end{figure*}

\begin{figure}
  \centering
        \includegraphics[width=\columnwidth]{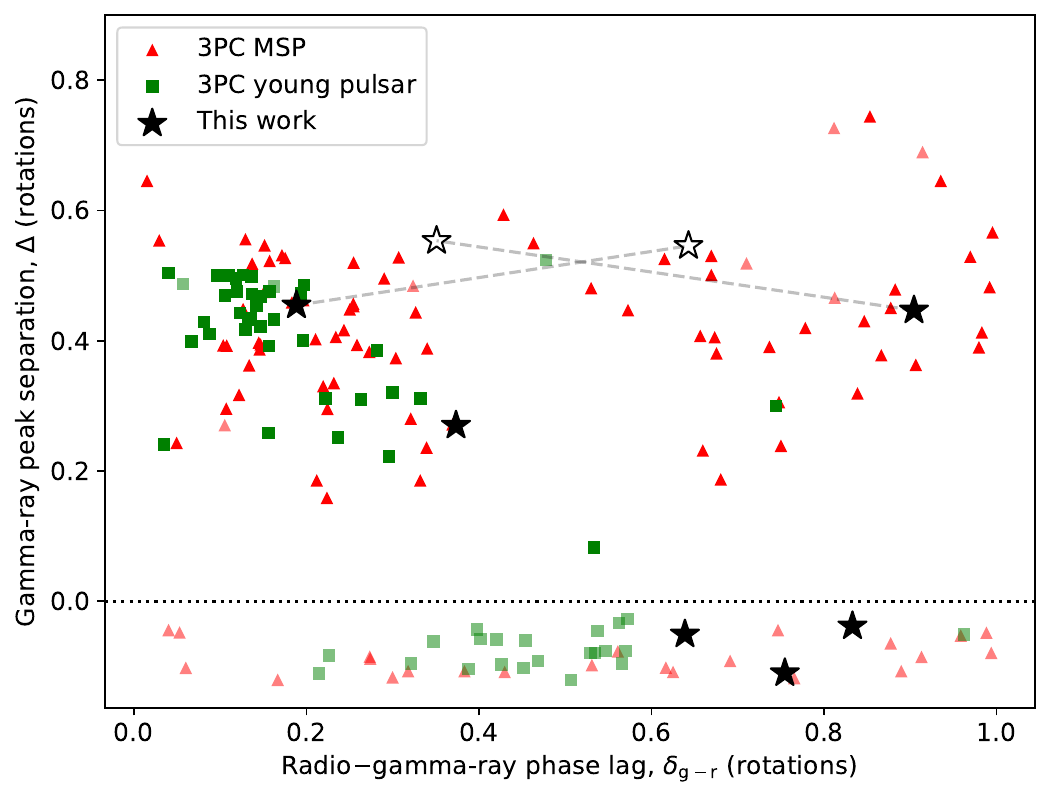}
        \caption{Gamma-ray peak separations ($\Delta$) vs. radio/gamma-ray phase-lag $\delta_{{\rm g}-{\rm r}}$ for gamma-ray pulsars from \citetalias{3PC} (coloured markers) and the MSPs studied here (black stars). For two pulsars, the definition of the primary gamma-ray peak is potentially ambiguous, and so for these we show the alternative values with empty markers, linked by dashed lines. Pulsars with only one gamma-ray peak are shown below $\Delta = 0$ with a random scatter in the y-axis for clarity.}
    \label{f:delta_Delta}
\end{figure}

A full modelling treatment is beyond the scope of this work, but we show the phase-aligned radio and gamma-ray pulse profiles for the 6 gamma-ray detected pulsars in Fig.~\ref{f:phase_alignment}, with the gamma-ray photon phases rotated to align with the fiducial zero-phase points of the radio pulse profiles. We also reproduce the $\delta_{{\rm g}-{\rm r}}$--$\Delta$ from the \citetalias{3PC} plot here in Fig.~\ref{f:delta_Delta}, along with $\delta_{{\rm g}-{\rm r}}$ and $\Delta$ (for pulsars with more than one gamma-ray peak in their pulse profile) for the MSPs studied here extracted from the gamma-ray pulse profile templates used for timing in Sect.~\ref{s:fermi}. We followed the same criteria used by \citetalias{3PC} to determine the first gamma-ray peak from which to measure $\delta_{{\rm g}-{\rm r}}$. For PSRs~J1526$-$2744 and J1858$-$5422 this determination is possibly ambiguous as these pulsars exhibit two peaks almost half a rotation apart; however, with our chosen definition, their $\delta_{{\rm g}-{\rm r}}$--$\Delta$ values lie within the bulk of the known population. Four of these MSPs have gamma-ray peaks that appear to arrive before the radio pulse ($\delta_{{\rm g}-{\rm r}} > 0.5$), a phenomenon often seen in MSPs but almost never in non-recycled pulsars.

\section{Polarimetric results}
\label{s:poln}

In this section, we briefly describe the main polarimetric properties of our MSPs. 
Stokes parameters have been recorded only at MeerKAT through the Pulsar Timing User Supplied Equipment (PTUSE; \citealt{Bailes2020+MeerTIME}) as the long integrations (hence, the large data size) were already prohibitive for the other telescopes. We therefore have polarimetry on all but two pulsars, J1526$-$2744 and J1803$-$6707, timed at Nan\c{c}ay and Parkes (MeerKAT data on these sources have been obtained, as part of the \textit{Fermi} L-band Shallow survey, with the Accelerated Pulsar Search User Supplied Equipment (APSUSE) recording total intensity information only). 

Polarisation calibration at MeerKAT is performed on-line as described in \cite{Serylak21}. All PTUSE data were then folded off-line using \texttt{DSPSR}\footnote{\url{http://dspsr.sourceforge.net/}} \citep{dspsr} and the best ephemeris for each pulsar. The pulsar software \texttt{PSRCHIVE}\footnote{\url{http://psrchive.sourceforge.net/}} \citep{psrchive} was used for all the subsequent steps: the routine \texttt{pac} was used to correct for parallactic angle variations and \texttt{pazi} was used to remove RFI. For each pulsar, cleaned, calibrated data were summed in phase with \texttt{psradd}, in order to obtain a high S/N profile. Finally \texttt{rmfit} was run to maximise the S/N of the linear polarisation component of the summed profiles and obtain a measurement of the rotation measure of each pulsar. 

\begin{figure}
  \centering
        \includegraphics[clip, trim=1.4cm 0.6cm 10cm 1cm, width=\columnwidth]{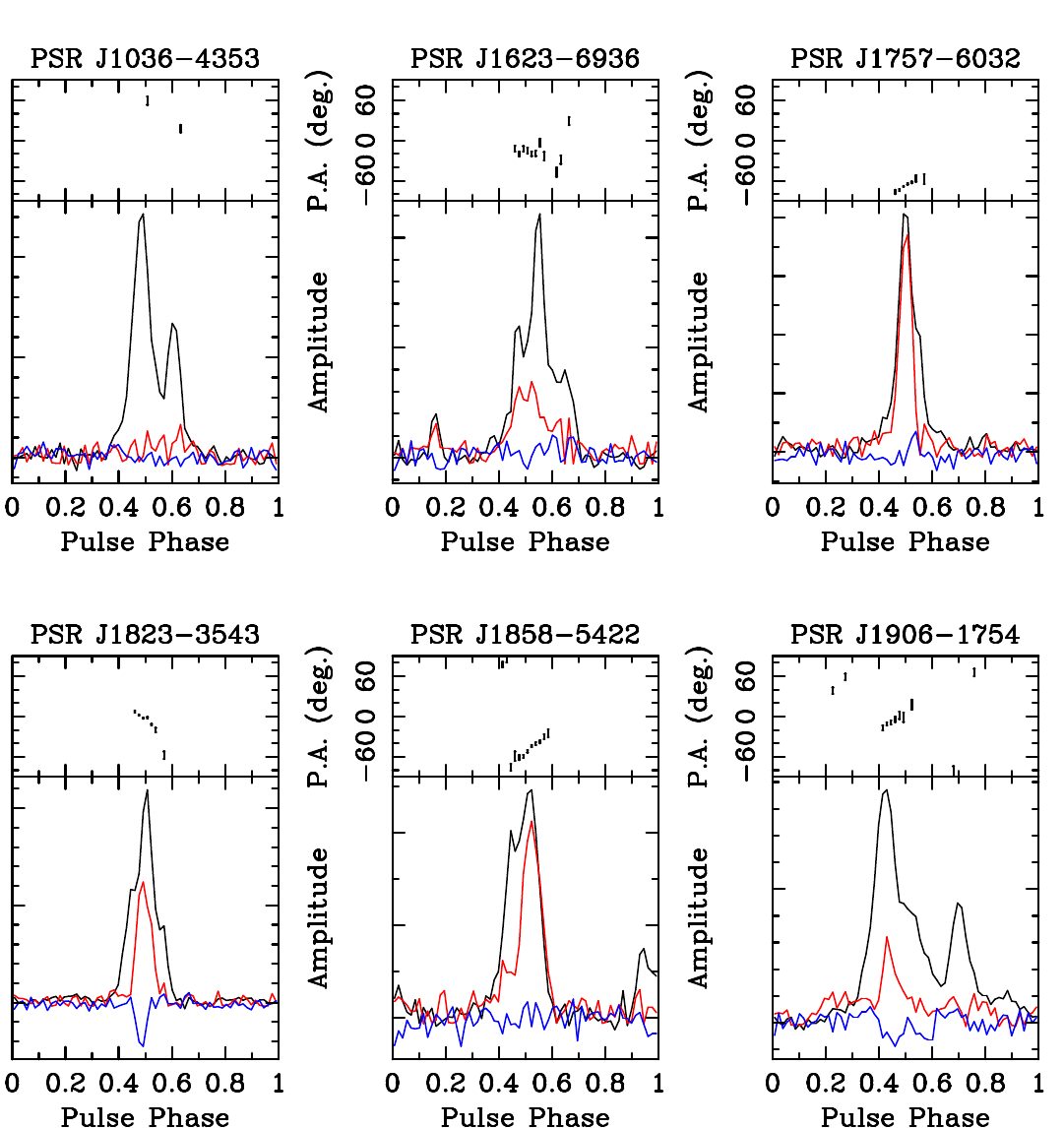}
        \caption{Polarisation profiles of the six \textit{Fermi} MSPs with
significant polarisation detections. The total intensity is shown in black,
while the total linear and circular polarisation are shown in red and blue,
respectively. The top panels for each pulsar show the polarisation position
angle (plotted for the profile bins where the linear polarisation has a S/N
larger than 3).}
    \label{f:poln}
\end{figure}

Figure \ref{f:poln} shows the calibrated, cleaned, de-rotated pulse profiles for the six pulsars (all except the isolated J1709$-$0333, which shows no detectable polarisation) obtained by summing all the observations collected with the MeerKAT L-band receiver (1.3 GHz). The black line is the total intensity profile, while red and blue are, respectively, the linear and circular polarisation profiles. The top part of each panel shows the variation of the polarisation position angle across the pulse rotational phase. 

Table \ref{tab:poln} lists, for the six pulsars of Fig. \ref{f:poln}, the percentage of linear (L) and circular (V) polarisation and its absolute value ($|$V$|$), and the rotation measure obtained from \texttt{rmfit}. 
The percentages of L,V, and $|$V$|$ were calculated following \cite{tiburzi+13}.

\begin{table}
  \centering
  \caption{Polarisation properties for six TRAPUM \textit{Fermi} MSPs.}
  \label{tab:poln}
  \begin{tabular}{lcccc}
    \hline
    Pulsar & \%L & \%V & \%$|$V$|$ & RM           \\
           &     &     &           & (rad m$^{-2}$)  \\
    \hline
    J1036$-$4353 & 11.1(13) & 1.7(12)  & 1.7(12) &$-$18(8)   \\ 
    J1623$-$6936 & 36.9(16) & 6.7(9)   & 7.7(9)  & 75(10)  \\
    J1757$-$6032 & 65.6(17) & $-$0.6(11) & 5.1(10) & 99(4)   \\
    J1823$-$3543 & 42.1(9)  & $-$7.9(8)  & 9.1(8)  & 62(4)   \\
    J1858$-$5422 & 67.3(22) & 0.0(17)  & 1.4(17) & 38(11)  \\
    J1906$-$1754 & 30.4(11) & $-$2.5(8)  & 7.3(8) & $-$26(12) \\
    \hline
  \end{tabular}
  \tablefoot{Left to right we report: pulsar name, percentage of linear polarisation, percentage of circular polarisation and of its absolute value, and rotation measure.}
\end{table}

\section{Summary and conclusions}
\label{s:summary}

In this paper, we present the results of a multi-telescope, multi-wavelength timing campaign on the nine MSPs (eight binaries and one isolated) discovered as part of the TRAPUM \textit{Fermi} Shallow survey at L-band \citep{clark+23}. Preliminary ephemerides obtained from radio observations allowed us to detect gamma-ray pulsations for six of our targets. Thanks to this, we were able to extend their timing solutions over the 15-year \textit{Fermi}-LAT data span. This was also possible for the RB pulsar J1803$-$6707, in spite of the presence of large orbital variations that are typical of RBs and that usually prevent the extrapolation of timing results beyond the radio data coverage; this was possible thanks to a new method (described in Sect. \ref{s:j1803} and initially given in \citealt{Thongmeearkom2024+RBs}). For the three pulsars for which no gamma-ray pulsed emission was found, a fully coherent timing solution was achieved with a typical longer radio follow-up campaign.

To take advantage of the full potential of both the radio and gamma-ray data, for the six sources with pulsations detected in both bands, we developed a new joint timing technique (described in Sect. \ref{s:radio-gamma}). It allowed us to obtain precise measurements for the long-term parameters constrained primarily by the gamma-ray photons and, in particular, to significantly measure the proper motion of four MSPs. The availability of well constrained, long- and short-term timing parameters is important not only to better characterise our targets, but also to extend their study at other wavelengths and through messengers other than the electromagnetic. Our timing results, for instance, are being currently used for a search of continuous gravitational waves over the entire LIGO data span (Ashok et al. in prep.).

The results of the (joint) timing campaign (presented in Sect. \ref{s:results}) show that our sample of pulsars differs from the previously known Galactic population (and, marginally, also from the earlier subset of Galactic MSPs discovered in association with \textit{Fermi} sources); these differences are mainly seen in the distribution of the spin periods, peaking at significantly lower values, in our case. We note that the difference in spin-period distribution does not correspond to a difference in spin-down energy distribution. It is therefore unlikely that the smaller spin periods of our targets are related to a difference in the selected gamma-ray targets. It may rather simply depend on the higher sensitivity of the MeerKAT telescope (allowing us to find fast-spinning sources also with larger duty cycles) and on the consequently shorter observing time required (allowing us to minimise the pulse smearing induced by the pulsar accelerating in a binary system). Shorter spin periods imply, potentially, smaller errors in the ToAs; shorter period pulsars have, therefore, a higher potential for being good candidates for high precision experiments. One example is the pulsar timing array experiments for the detection of nano-hertz gravitational waves, where MeerKAT promises to be a powerful asset.

Individual sources of interest reported in this paper include the two eclipsing RBs J1036$-$4353 and J1803$-$6708, whose timing results are being used to complement optical observations (Phosrisom et al. in prep.) and will allow us to constrain the masses of the NSs in the systems, adding valuable insights into the NS mass distribution and potentially supplementing our understanding of the EoS of nuclear matter. Similarly, albeit through radio timing-only, the high-mass binary containing PSR J1757$-$6032, has also demonstrated potential for precise NS mass measurements through the detection (currently with limited significance) of the Shapiro delay relativistic effect. This pulsar appears to be one of the few known objects to have been formed through a Case A Roche lobe overflow: an evolutionary path possibly leading to the formation of  massive NSs. Further dedicated observations will be needed to fully characterise this system.

Building on our results and on the methods developed to fully exploit the potential of our multi-wavelength data set, we have now expanded our searches of \textit{Fermi} targets both in terms of frequency (observing the same sources covered in our L-band survey also at 800 MHz; Thongmeearkom et al. in prep.) and source sample (both new gamma-ray-only-selected targets and spider candidates selected through multi-wavelength information; \citealt{Thongmeearkom2024+RBs}). The preliminary results once again confirm the high scientific potential of pulsar experiments targeting \textit{Fermi} unidentified point-like sources, especially when coupled to such a sensitive and versatile instrument as the MeerKAT telescope.

\section*{Data availability}
\label{sec:dataAvail}
The full ORBIFUNC and ORBWAVES ephemeris for PSR J1803$-$6708 can be downloaded from Zenodo at the following link: \url{https://zenodo.org/records/13862732}.

\begin{acknowledgements}
The MeerKAT telescope is operated by the South African Radio Astronomy Observatory, which is a facility of the National Research Foundation, an agency of the Department of Science and Innovation. SARAO acknowledges the ongoing advice and calibration of GPS systems by the National Metrology Institute of South Africa (NMISA) and the time space reference systems department of the Paris Observatory. TRAPUM observations use the FBFUSE and APSUSE computing clusters for data acquisition, storage and analysis. These clusters were funded, installed and operated by the Max-Planck-Institut f\"{u}r Radioastronomie and the Max-Planck-Gesellschaft. PTUSE was developed with support from the Australian SKA Office and Swinburne University of Technology.
Murriyang, the Parkes radio telescope, is part of the Australia Telescope National Facility (\url{https://ror.org/05qajvd42}) which is funded by the Australian Government for operation as a National Facility managed by CSIRO. We acknowledge the Wiradjuri people as the traditional owners of the Observatory site. 
This publication is partly based on observations with the 100-m telescope of the MPIfR (Max-Planck-Institut für Radioastronomie) at Effelsberg.
The Nan\c{c}ay Radio Observatory is operated by the Paris Observatory, associated with the French Centre National de la Recherche Scientifique
(CNRS) and Université d’Orléans. It is partially supported by the Region Centre Val de Loire in France.
This work was supported in part by the “Italian Ministry of Foreign Affairs and International Cooperation”, grant number ZA23GR03, under the project "RADIOMAP-Science and technology pathways to MeerKAT+: the Italian and South African synergy"

MBu and AP acknowledge the use of resources from the research grant "iPeska" (P.I. Andrea Possenti) funded under the INAF national call Prin-SKA/CTA (approved with the Presidential Decree 70/2016), and the use of the TRG computer cluster at INAF - Cagliari, funded by the Autonomous Region of Sardinia (Regional Law 7 August 2007 n. 7, year 2015, “Highly qualified human capital”; P.I. Marta Burgay). 
R.P.B. acknowledges support from the European Research Council (ERC)
under the European Union's Horizon 2020 research and innovation programme (grant agreement No. 715051; Spiders)

The National Radio Astronomy Observatory is a facility of the National Science Foundation operated under cooperative agreement by Associated Universities, Inc. SMR is a CIFAR Fellow and is supported by the NSF Physics Frontiers Center award 2020265.

The \textit{Fermi}-LAT Collaboration acknowledges generous ongoing support from a number of agencies and institutes that have supported both the development and the operation of the  LAT  as  well  as  scientific  data  analysis.  These  include  the National  Aeronautics  and Space   Administration   and   the   Department   of   Energy   in the   United   States,   the Commissariat \`{a} l'Energie Atomique and the Centre National de la Recherche Scientifique /  Institut  National  de  Physique  Nucl\'{e}aire et  de  Physique  des  Particules  in  France,  the Agenzia  Spaziale  Italiana and the Istituto  Nazionale  di  Fisica  Nucleare  in  Italy,  the Ministry  of  Education,  Culture, Sports,  Science  and  Technology  (MEXT),  High  Energy Accelerator  Research Organization  (KEK)  and  Japan  Aerospace  Exploration  Agency (JAXA)  in  Japan, and  the  K.  A.  Wallenberg  Foundation,  the  Swedish  Research  Council and the Swedish National Space Board in Sweden. 

Additional   support   for   science   analysis   during   the   operations phase is gratefully acknowledged from the Istituto Nazionale di Astrofisica in Italy and the Centre National d'Etudes Spatiales  in  France. This  work  performed  in  part  under  DOE  Contract  DE-AC02-76SF00515.

We gratefully acknowledge Thankful H. Cromartie, Matthew Kerr, David A. Smith and David J. Thompson for reviewing this manuscript on behalf of the \textit{Fermi}-LAT collaboration. 
 \end{acknowledgements}

\bibliographystyle{aa}
\bibliography{ms}

\onecolumn
\begin{appendix}
\section{Additional figures}
\label{s:appendix}
\begin{figure}[!hb]
  \centering
        \includegraphics[width=\textwidth]{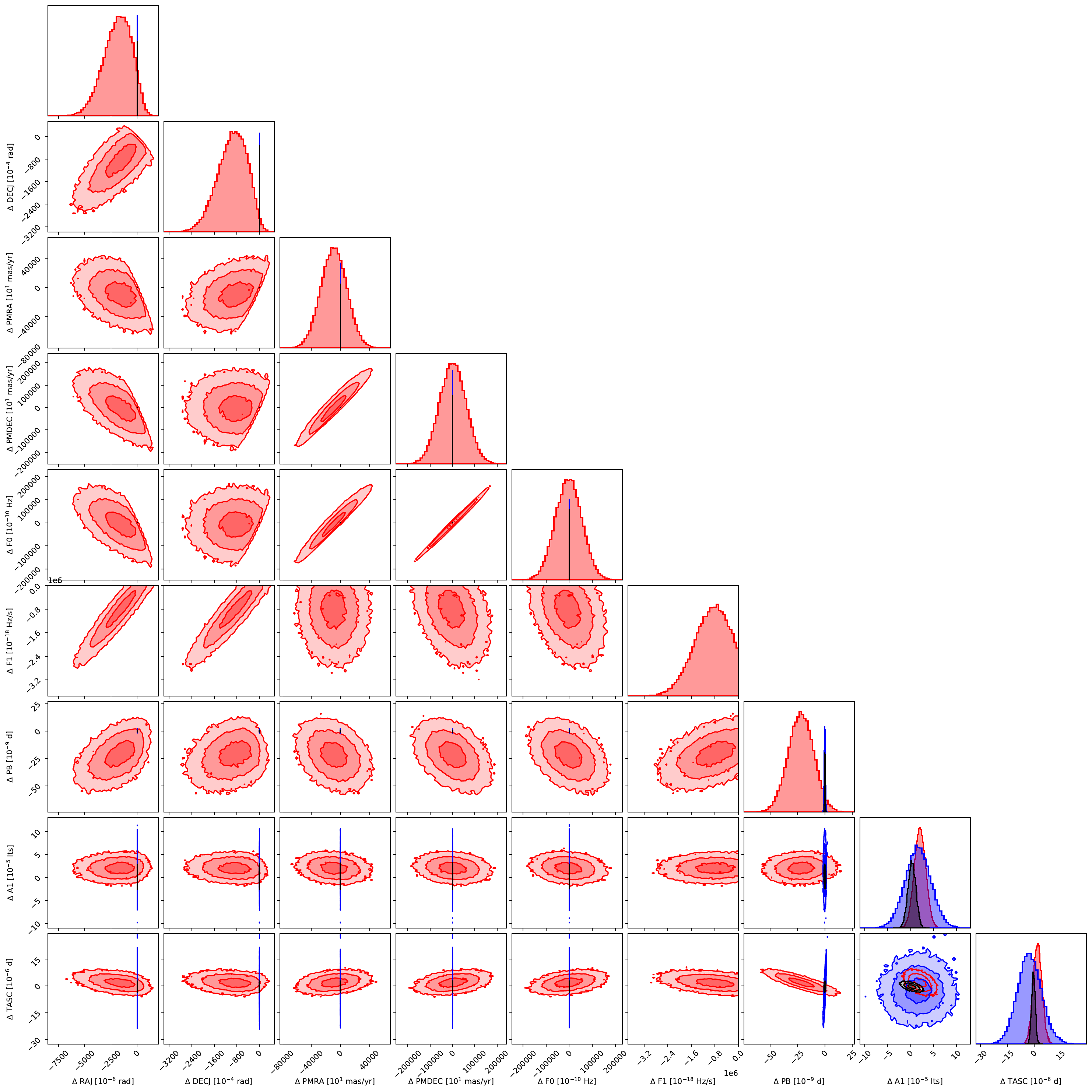}
        \caption{Corner plot for all the jointly fitted parameters for J1526-2744 (colours as in Fig. \ref{f:j1858_astro}).}
    \label{f:j1526_compare}
\end{figure}
\begin{figure}
  \centering
        \includegraphics[width=\textwidth]{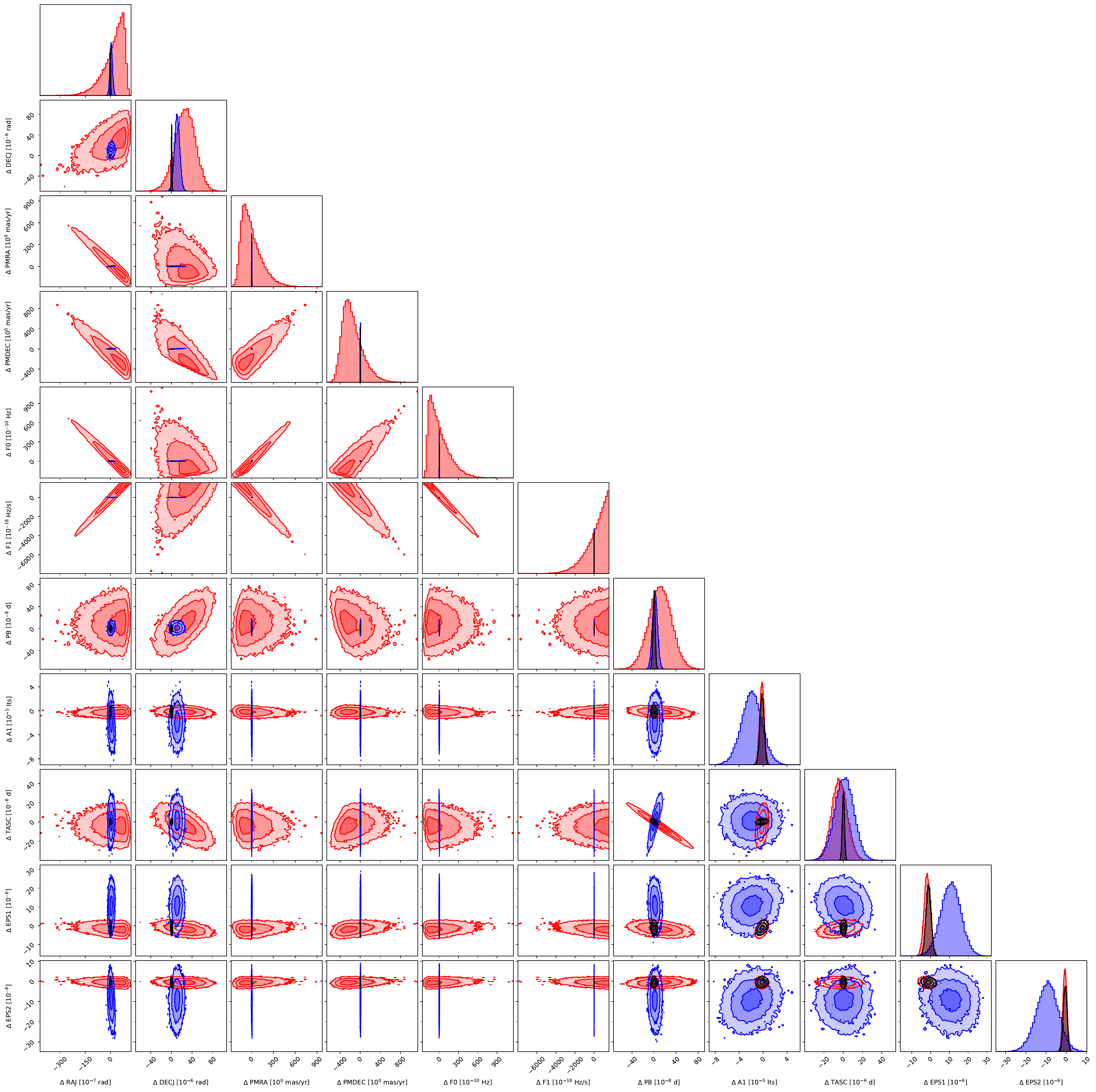}
        \caption{Corner plot for all the jointly fitted parameters for J1623-6936 (colours as in Fig. \ref{f:j1858_astro}).}
    \label{f:j1623_compare}
\end{figure}
\begin{figure}
  \centering
        \includegraphics[width=\textwidth]{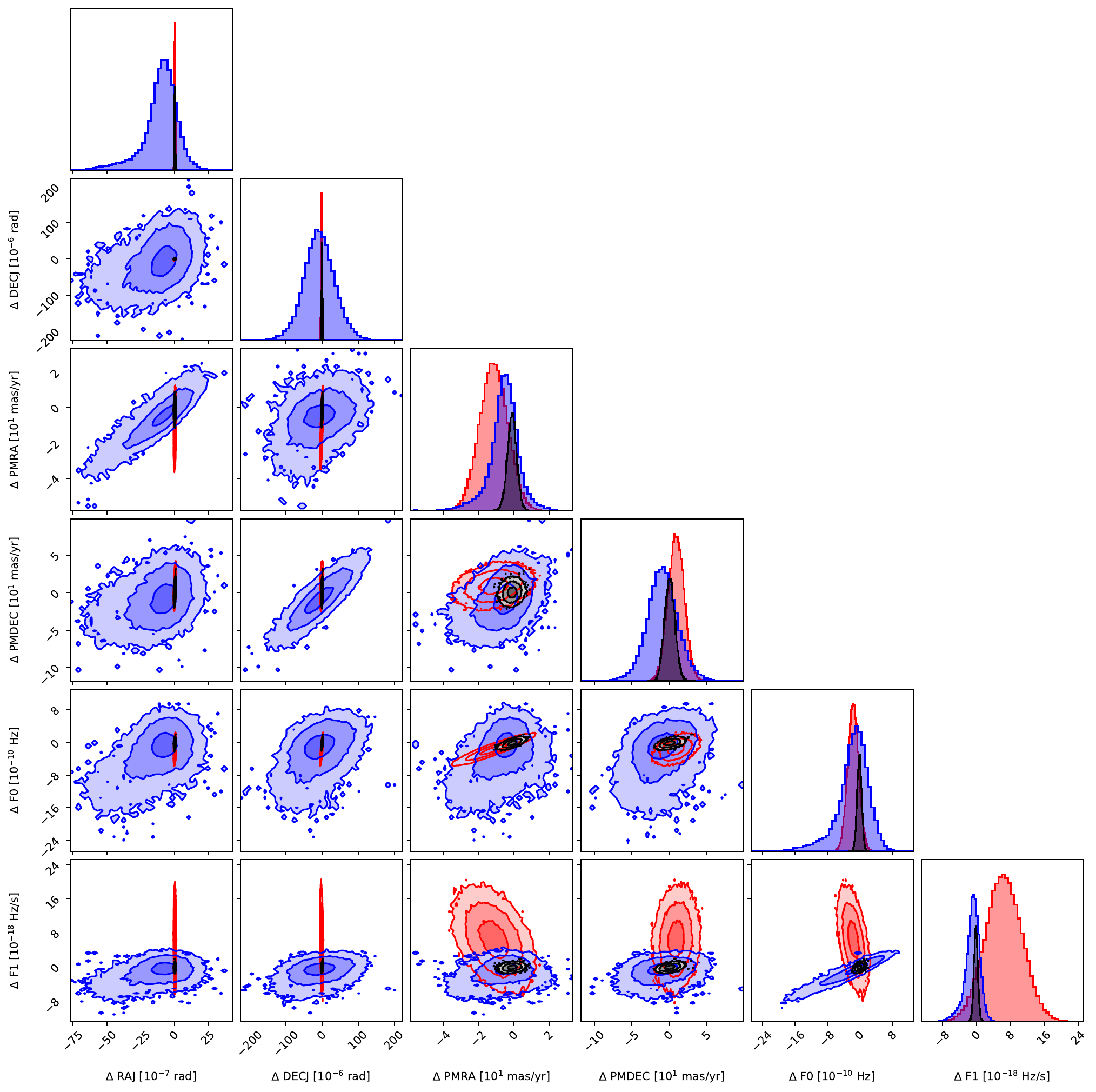}
        \caption{Corner plot for all the jointly fitted parameters for J1709-0333 (colours as in Fig. \ref{f:j1858_astro}).}
    \label{f:j1709_compare}
\end{figure}
\begin{figure}
  \centering
        \includegraphics[width=\textwidth]{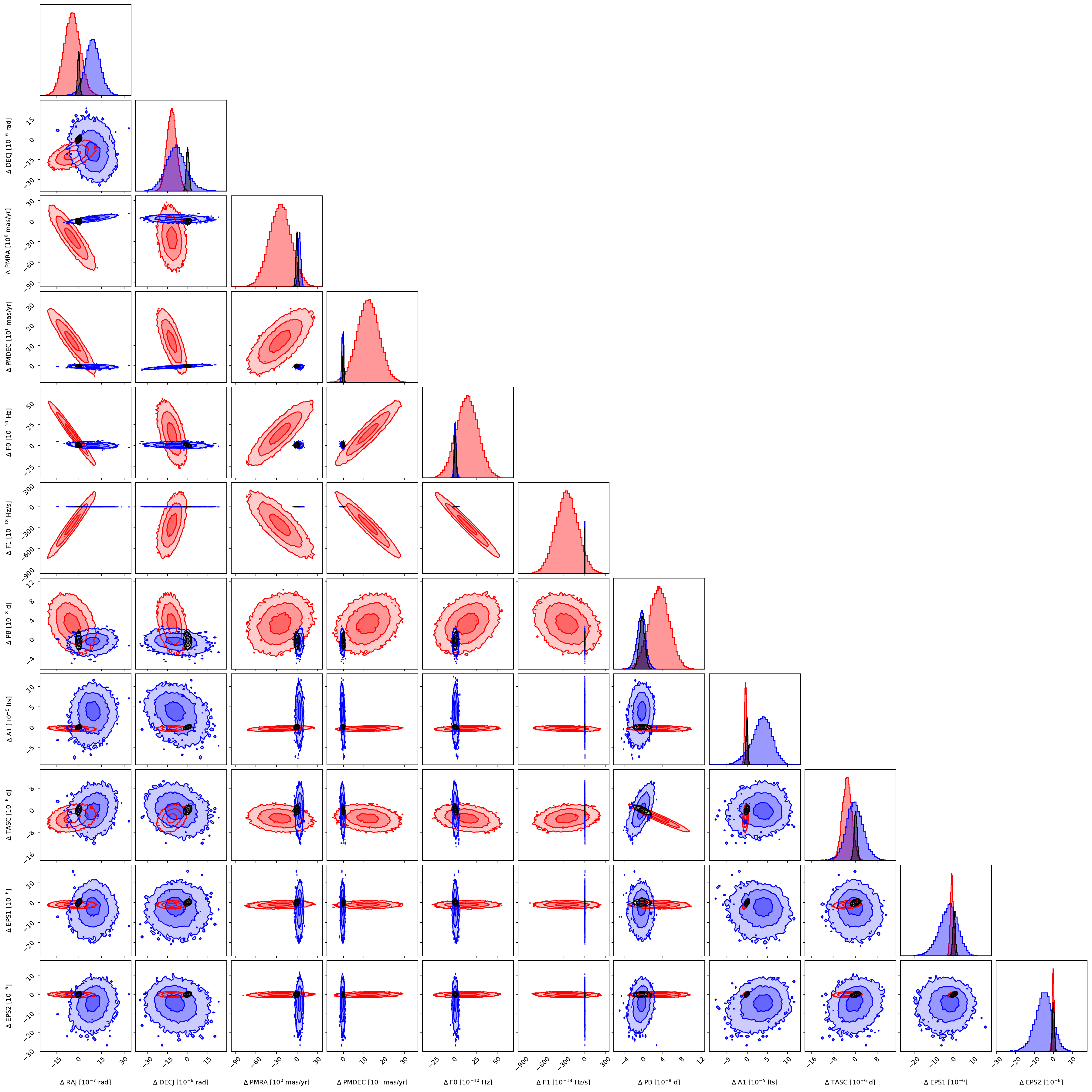}
        \caption{Corner plot for all the jointly fitted parameters for J1757-6032 (colours as in Fig. \ref{f:j1858_astro}).}
    \label{f:j1757_compare}
\end{figure}
\begin{figure}
  \centering
        \includegraphics[width=\textwidth]{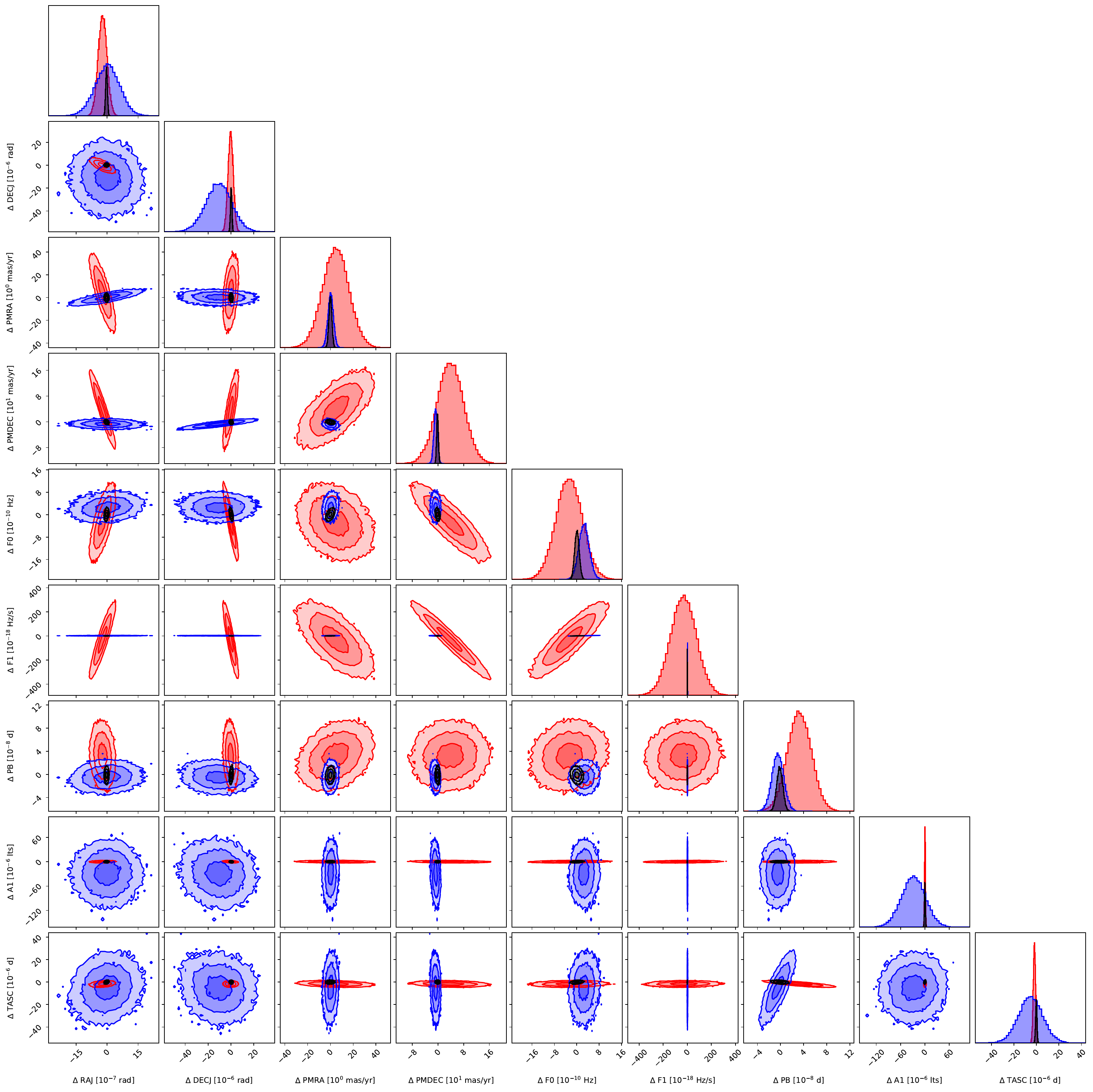}
        \caption{Corner plot for all the jointly fitted parameters for J1858-5422 (colours as in Fig. \ref{f:j1858_astro}).}
    \label{f:j1858_compare}
\end{figure}
\end{appendix}

\end{document}